\documentclass[fleqn,usenatbib]{mnras}

\usepackage{newtxtext, newtxmath}

\usepackage[T1]{fontenc}

\DeclareRobustCommand{\VAN}[3]{#2}
\let\VANthebibliography\thebibliography
\def\thebibliography{\DeclareRobustCommand{\VAN}[3]{##3}\VANthebibliography}

\usepackage{graphicx}	
\usepackage{amsmath}	
\usepackage{subfig}     
\usepackage{makecell}   
\usepackage{url}

\defcitealias{Collett2018}{C18}

\title[Probe of $\eta_{\text{PPN}}$ at galaxy-scale]{Probing General Relativity in galactic scales at z $\sim0.3$}

\author[C. R. Melo-Carneiro et al.]{
Carlos R. Melo-Carneiro,$^{1}$\thanks{E-mail: carlos.melo@ufrgs.br},
Cristina Furlanetto$^{1}$,
Ana L. Chies-Santos$^{1,2}$
\\
$^{1}$Instituto de F\'isica, Universidade Federal do Rio Grande do Sul, Porto Alegre, RS, Brazil\\
 $^{2}$Shanghai Astronomical Observatory, Chinese Academy of Sciences, 80 Nandan Rd., Shanghai 200030, China\\
}

\date{Accepted XXX. Received YYY; in original form ZZZ}

\pubyear{2015}

\begin{document}

\label{firstpage}
\pagerange{\pageref{firstpage}--\pageref{lastpage}}
\maketitle

\begin{abstract}
General Relativity (GR) has been successfully tested mainly at Solar system scales; however, galaxy-scale tests have become popular in the last few decades. In this work, we investigate the $\eta_\text{PPN}$ parameter, which is commonly defined by the ratio of two scalar potentials that appears in the cosmological linearly perturbed metric. Under the assumption of GR and a vanish anisotropic stress tensor, $\eta_\text{PPN}= 1$. Using ALMA, HST, and VLT/MUSE data, we combine mass measurements, using gravitational lensing and galactic dynamics, for the SDP.81 lens galaxy ($z = 0.299$) to constrain $\eta_\text{PPN}$. By using a flexible and self-consistent mass profile, our fiducial model takes into account the contribution of the stellar mass and a dark matter halo to reconstruct the lensed galaxy and the spatially-resolved stellar kinematics. We infer, after accounting for systematic uncertainties related to the mass model, cosmology and kinematics, $\eta_{\text{PPN}} = 1.13^{+0.03}_{-0.03}\pm0.20\,(\text{sys})$, which is in accordance with GR predictions. Better spectroscopy data are needed to push the systematics down and bring the uncertainty to the percentage level since our analysis shows that the main source of the systematics is related to kinematics, which heavily depends on the signal-to-noise ratio of the spectra. 
\end{abstract}




\section{Introduction}\label{Sec. Intro}

General Relativity (GR) is one of the most successful theories in Physics to date and has been tested in different scenarios since its publication in 1915 \citep[e.g.][]{Dyson1920, Akiyama2019}. Over the last decades, more precise tests have been performed in order to push the limits of the theory, in particular on the solar system scales \citep[e.g.][]{Bertotti2003, Lunar_Laser2004}. However, few tests were performed on galactic and cosmological scales, and only in recent years more effort has been spent in this direction.

The current standard model for cosmology, $\Lambda$-Cold Dark Matter ($\Lambda$CDM), which is based on GR, requires two components not directly observed yet, dark matter and dark energy. Although such model represents the simplest explanation for the lack of visible matter in the Universe and the accelerated cosmological expansion, the nature of these components is still poorly understood. Perhaps more critically, some theoretical questions remain open about the phenomenology of the dark components \citep[][]{Salucci2019, Motta2021}. In addition, the so-called  Hubble tension \citep{Riess_2019} could indicate another problem with our knowledge about the cosmological model or even about our knowledge on the gravitational interaction.

In order to understand these tensions, some alternatives, such as modifying the equation of state of dark energy \citep[e.g.][]{2019ApJ...883L...3L,2020A&A...641A...6P}, unified models of dark matter and dark energy  \citep[e.g.][]{PhysRevD.88.063519, 2020PhRvD.101l3513G}, and even modifications on Einstein's theory of gravity, have been proposed so far \citep[e.g.][]{2016PhRvD..93b3513D, 2019ApJ...886L...6S, Ishak2019}.

Considering the Newtonian limit and small scales, i.e., sub-horizon length scales where the Hubble Flow can be considered near-constant, a common approach is to use the Parametrized Post-Newtonian (PPN) formalism \citep{Will2014, Will2018}, a well-known framework for testing the weak-field regime of gravity. In particular, the PPN parameter, $\gamma_{\text{PPN}}$, has received a lot of attention over the last decades \citep[e.g.][]{Bertotti2003,Schwab2010,Cao2017}. Under the assumption of GR, $\gamma_{\text{PPN}} = 1$. Thus, any deviation from this value would indicate a possible violation of the current theory of gravity.

On the other hand, when considering cosmological scales, i.e., horizon length scales where the Hubble Flow is no longer constant over the entire domain considered, tests involving the linearized gravity \citep{Carroll2014, Mo2010} are more common. In this work, we focus our attention on the gravitational slip parameter $\eta$, which is defined here by the ratio of two scalar potentials that appear in the linear perturbed Friedmann-Lemaître-Robertson-Walker (FLRW) metric,

\begin{equation}\label{eq: line element ppn}
    dS^2 =  -\left(1+2\frac{\Phi}{c^2}\right)c^2 dt^2 + \left(1-2\frac{\Psi}{c^2}\right)h_{ij}dx^idx^j,
\end{equation}
where $c$ is the speed of light, $t$ is the time, $h_{ij}$ is the three-metric tensor of constant curvature space, and $dx^i$ are spatial coordinates. 

The first potential $\Phi$ is the classical Newtonian potential,  more important to the motion of non-relativistic particles ($v^2/c^2 << 1$). The second potential $\Psi$, oppositely, is more relevant to the motion of relativistic particles ($v^2/c^2 \sim 1$), and since it is associated with the spatial curvature of the metric, it is called curvature potential. The interpretation of these potentials can be understood as stated by \citet{Simpson_2012}: `\textit{is the strength of gravity the same on cosmological scales as it is here on Earth?}'. If the answer is no, then maybe, the motion of non-relativistic and relativistic particles can be modified differently due to the presence of these two potentials.

In this scenario, we define the gravitational slip parameter $\eta$, as 
\begin{equation}\label{eq: slip parameter}
    \eta= \frac{\Psi}{\Phi},
\end{equation}
which can be understood as an effective gravitational coupling between light and matter since the potentials, in principle, can act differently on both components. The $\eta$ parameter can be a function of time and scale \citep{Ma1995, Bertschinger2008}, and also can assume different values in different gravitational theories. Assuming GR, with a vanishing anisotropic stress tensor\footnote{For violations of this condition  see, e.g. \citet{Jain_2008}.}, $\Phi = \Psi$, such that $\eta = 1$. Therefore, a deviation of this parameter from the unit, once again, could indicate a violation of the standard gravitational model based on GR or a problem with our knowledge of the cosmological model.

Although $\eta$ and $\gamma_{\text{PPN}}$ have the same numerical value, they are not the same or even have the same observational constraints in general, as shown by \citet{Toniato_2021}. Even so, under certain considerations, the $\gamma_{\text{PPN}}$ and $\eta$ parameters can be connected such that we are able to impose bounds on both using the same observational constraints. To do so, we assume the following: 
\begin{enumerate}
        \item The space-time metric is given by the line element equation~(\ref{eq: line element ppn}), which is in the Newtonian gauge and considers only scalar perturbations;
        \item There is a well-defined Newtonian limit, where the potentials $\Phi$ and $\Psi$ still follow the Poisson equation;
        \item The gravitational slip parameter is constant on the relevant scales being studied;
\end{enumerate}
Under these considerations, $\gamma_{\text{PPN}} = \eta$, such that it is possible to find bounds for both parameters using the same observable constraints \citep[][]{Toniato_2021}. Moreover, to avoid possible confusion between the general definition of the slip parameter, equation~(\ref{eq: slip parameter}), and the $\eta$ defined under the assumptions presented above, we will call the latter definition by $\eta_{\text{PPN}}$, and the more general definition just by $\eta$, as done by \citet{Toniato_2021}.

Many of the recent tests of the $\gamma_{\text{PPN}}$, outside the solar system, are concentrated in the light deflection provided by the strong gravitational lensing (SGL), which provides a straightforward way to constrain it at galactic scales.

Using measurements of velocity dispersion and gravitational lensing \citet{Schwab2010} found $\gamma_{\text{PPN}} = 1.01 \pm {0.05}$ for a sample of 53 galaxies with  $z \sim 0.1 - 0.3$. Following the same methodology, \citet{Cao2017} extends the sample size of \citeauthor{Schwab2010} to 80 lens galaxies in the redshift range $z \sim 0.08 - 0.94$, thereby determining a value of  $\gamma_{\text{PPN}} = 0.995^{+0.037}_{-0.047}$. Both inferences are in agreement with GR, although the estimated systematic uncertainties (around $25\%$ for the second sample), in lens modelling and velocity dispersion measurements are dominant in this kind of study.

Nevertheless, not only the PPN parameter can be constrained by SGL data, but also the slip gravitational parameter as well. Due to the nature of $\eta$, the motion of non-relativistic and relativistic particles can be differentially affected, such that we can probe it by comparing the mass inferred by different tracers (e.g. cluster dynamics and gravitational lensing), which are sensitive to the different potentials of the same extragalactic object. Based on this approach, and using data from the galaxy cluster MACS J1206.2-0847 at $z = 0.44$, \citet{Pizzuti2016} probed the slip parameter applying galaxy cluster dynamics and gravitational lensing phenomena. They found a value of $\eta(r_{200}) = 1.01^{+0.31}_{-0.28}$ consistent with GR, where $r_{200}$ indicates that their inference was made for a fixed radius of a sphere with density 200 times the critical density of the Universe at that redshift\footnote{Note that, in this work, the authors did not make any of the assumptions that we presented above.}.

Similarly, considering the same assumptions as those previously presented here, \citet{Collett2018}, hereafter \citetalias{Collett2018}, imposed the most precise constraint on $\eta_{\text{PPN}}$ at galactic scales in the literature to date, which can be also  interpreted as a constraint of $\gamma_{\text{PPN}}$. Combining high-resolution lensing data with kinematical data inferred from integral field spectroscopy (IFS)  for a system at redshift $z = 0.035$, \citetalias{Collett2018} found a value of $\eta_{\text{PPN}} = 0.97 \pm 0.09$, including the systematic uncertainty, which strongly agrees with GR. 

In this work, we propose to probe the $\eta_{\text{PPN}}$, using the gravitational lens system H-ATLAS J090311.6+003906 \citep[SDP81;][]{Negrello2014} at intermediate redshift ($z = 0.299$), employing consistent methodology as applied in \citetalias{Collett2018}, although with additional data coming from interferometry observations for the lensing part of the analysis. The $\eta_{\text{PPN}}$ parameter is probed by modelling lensing and dynamical masses simultaneously in a self-consistent way. These masses, in turn, are related by

\begin{equation}\label{eq: PPN masses}
    M_\text{dyn} = \frac{1 + \eta_{\text{PPN}}}{2}M_{\text{lens}}^{\text{GR}},
\end{equation}
where $M_\text{dyn}$ is the mass inferred by the kinematic modelling, and $M_{\text{lens}}^{\text{GR}}$ the mass inferred by the lens modelling assuming GR (\citetalias{Collett2018}).

This paper is organised as follows. In Section~\ref{Sec. Data} we describe the lens system and the data used for our analysis. Section~\ref{Sec. Mass Profile} describes the mass model adopted and how we parametrize it. In Section~\ref{Sec. Modelling} we describe the modelling procedure and how the joint analysis is performed. Sections~\ref{Sec. Results} and~ \ref{Sec. Discussion} are intended to present and discuss the fiducial model, and in Section~\ref{Sec. Conclusions} we present a summary and the final thoughts.

Throughout the paper, unless otherwise stated, the cosmological parameters assumed are: $H_0 = 67.7$\,km\,s$^{-1}$\,Mpc$^{-1}$, $\Omega_{\Lambda} = 0.6911$, $\Omega_{m} = 0.3089$ \citep{Planck_2015}.
\section{Data}\label{Sec. Data}
The SDP.81 system, composed by an early-type foreground galaxy at $z_l = 0.299$ and a submm source galaxy at $z_s=3.042$, was first detected by \citet{Negrello2010} as part of the Herschel Astrophysical Terahertz Large Area Survey (H-ATLAS). The system was extensively studied  \citep[e.g.][]{Negrello2014, Dye2015, Tamura2015, Rybak2015, Wong2015}, with the main focus on the reconstruction of the source galaxy and on its properties, motivated by the unprecedented tens-of-parsec resolution of its submillimeter data from the  Atacama Large Millimeter/Submillimeter Array (ALMA) observatory. 

To impose constraints on $\eta_{\text{PPN}}$, combining different datasets that provide complementary information about the same massive object is necessary. The dataset used in this work is composed of high-resolution photometric data from the Hubble Space Telescope (HST), IFS data from the Multi-Unit Spectroscopic Explorer (MUSE), and interferometric data from ALMA. We present an overview of these datasets in the following, although more details can be found in the related papers.

\subsection{HST data}
The HST data were obtained with the Wide Field Camera 3 (WFC3) in 2011 in two different bands: F160W and F110W. Both images are publicly available in the Hubble Legacy Archive\footnote{\url{https://hla.stsci.edu/}} (PropID: 12194, PI: Negrello). The data were reduced using the WFC3 standard pipeline by the Barbara A. Mikulski Archive for Space Telescopes (MAST) team, and drizzled following \citet{2012drzp.book.....G}.

For our purposes, we used the deepest image, F160W, which has a pixel scale of $0.09\arcsec$. The point spread function (PSF) was obtained fitting a collection of unsaturated stars in  the field near the galaxy using the \texttt{IRAF} task \texttt{DAOPHOT} \citep{DAOPHOT}, resulting in a PSF with full width at half maximum (FWHM) approximately equal to $0.108\arcsec $, which corresponds to $\sim 495$\,pc at the redshift of the lens galaxy.

The lens galaxy light profile was fitted as a sum of 2D-elliptical Gaussians, which is used as a tracer of the stellar mass profile, up to the stellar mass-to-light ratio ($M/L$) factor; details are given in Section~\ref{Sec. Mass Profile}. After modelling the lens galaxy light profile and subtracting it from the original image, the deflected source is revealed. However, the lens-subtracted image is not used for modelling since better interferometric data are available.

\begin{figure*}
	\includegraphics[width=\textwidth]{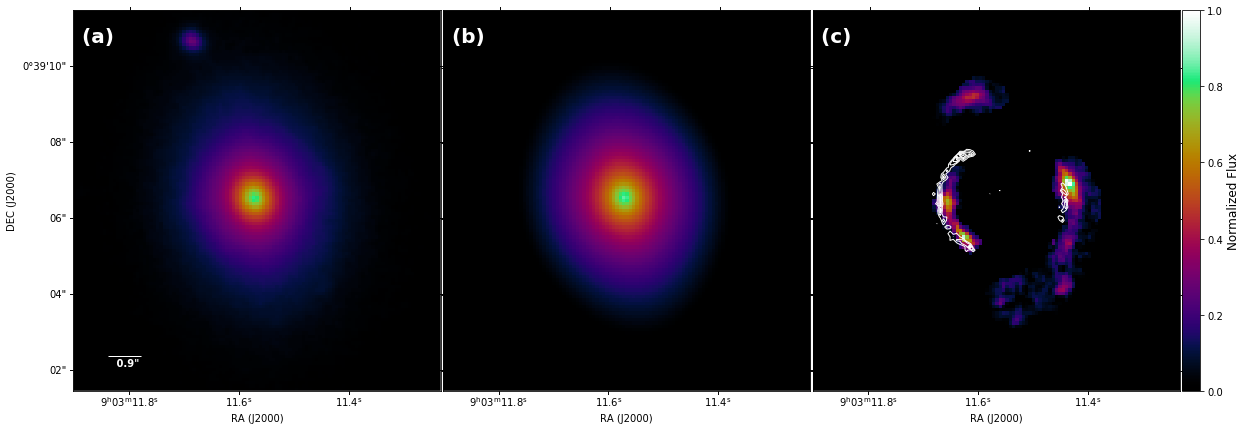}
    \caption{Panel (a) Interpolated HST/WCF3-F160W image. In the top left side we see the emission of a star, which is masked for the fit. Panel (b) Final MGE model after the interpolation in the position of the arcs. Panel (c) Lensed source after subtraction of the foreground lens light with overlaid contours of ALMA band 7 emission, see Sec.~\ref{sub:Alma-data} for details. Note that the residuals of the lens light are masked. All three panels have the same colour bar. North is up, and East is left.}
    \label{fig:hst-data}
\end{figure*}

To mitigate possible contamination due to the emission of the lensed source in the lens surface brightness profile fitted, we subtract the arcs from the original image. We interpolate the subtracted image across the regions where the arcs were excluded (see panel (c) of Figure~\ref{fig:hst-data}), obtaining a lens image without the source emission. Then a new galaxy light profile is fitted using the interpolated image, resulting in a final model for the surface brightness profile, which is used to trace the stellar mass. The HST interpolated image, the final model and the residuals showing the deflected source can be seen in Figure~\ref{fig:hst-data}, with the ALMA emission overlaid in panel (c).

\subsection{MUSE data}
SDP.81 MUSE data were obtained from the ESO Science Archive Facility\footnote{\url{http://archive.eso.org/cms.html}} (ProgID: 294.B-5042, PI: Gavazzi), with $13600$\,s of total exposure time. The data cover the spectral range of $460-935$\,nm, with mean spectral Resolution (R) equal to 2989 and spectral scale of $1.25$\AA.  The field of view is $1.64\arcmin \times 1.64\arcmin$ with a pixel scale of $0.2\arcsec $. From the collapsed image, a PSF (FWHM$=0.421\arcsec $, which corresponds to  $\sim 1930$\,pc at the redshift of the lens galaxy) was built following the same procedure used for the HST image.  To align the MUSE data with the HST data, we use the HST image as a reference and perform an astrometric shift equal to $0.18\arcsec $ in the MUSE data using the \texttt{ASTROALIGN} \citep{beroiz2019astroalign} package.

The MUSE data were reduced as a product of the ESO Phase 3 archive, following the standard pipeline described by \citet{2016ascl.soft10004W}. We use the  Zurich Atmosphere Purge code \citep[\texttt{ZAP};][]{ZAP} to remove telluric lines which may have been left in the previous reduction step.

\begin{figure}
	\includegraphics[width=\columnwidth]{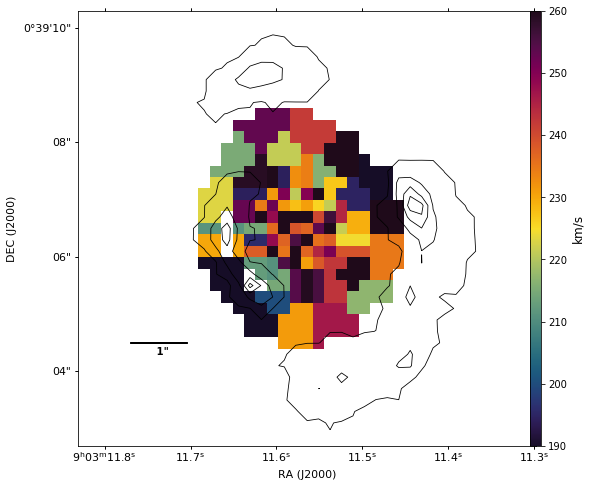}
    \caption{Resulting $V_{\text{rms}}$ map of SDP.81 lens galaxy. All the spaxels are shown here, not only the Voronoi binned ones. Overlaid contours of HST/WFC3-F160W source emission. The blank areas are the spaxels removed due to the strong contamination of the emission from the lensed source. North is up, and East is left.
}
    \label{fig:muse-data}
\end{figure}

In order to obtain the kinematical information used to constrain the dynamical mass (see Section~\ref{Sec. Modelling}), we model the spectral data obtained with MUSE using the \texttt{pPXF} package \citep{Cappellari_2004, Cappellari2016_PPXF}. We only select spectra with a signal-to-noise ratio (SNR) above $2.0$, to improve the spectral modelling. The SNR is measured as the ratio of the average signal and the average noise in the wavelength range $5000 - 7000$\,\AA\, (in the galaxy frame), the same range used for the spectral modelling. To minimise the possible contamination of the lensed source, we identify the emission peaks of the source in the HST image. After overlaying it on the MUSE collapsed image, we remove the spaxels of the MUSE data that correspond to those pixels with a strong source emission (roughly responsible for a $40$\% of the total emission on the East arc, see Figure~\ref{fig:muse-data}). Then, to ensure a reliable stellar kinematics measurement, we sum nearby spectra using the adaptive spatial two-dimensional binning scheme from \texttt{VorBin} package \citep{Vorbin_2003} to increase the SNR in each resulting Voronoi bin at a minimum of 10.

Each Voronoi bin is modelled using \texttt{pPXF} as a combination of a select set\footnote{Available at \url{https://github.com/remingtonsexton/BADASS3}.} of templates from the Indo-US stellar library\footnote{\url{https://www.noao.edu/cflib/}} \citep{Indo-US} in the optical region $3460-9464$\,\AA.  We choose the Indo-US templates due to its spectral resolution of FWHM $=1.35$\,\AA, which is appropriate for modelling spectra at the redshift of SDP.81, allowing us to convolve the templates with the MUSE instrumental FWHM.

\begin{figure}
	\includegraphics[width=1.05\columnwidth]{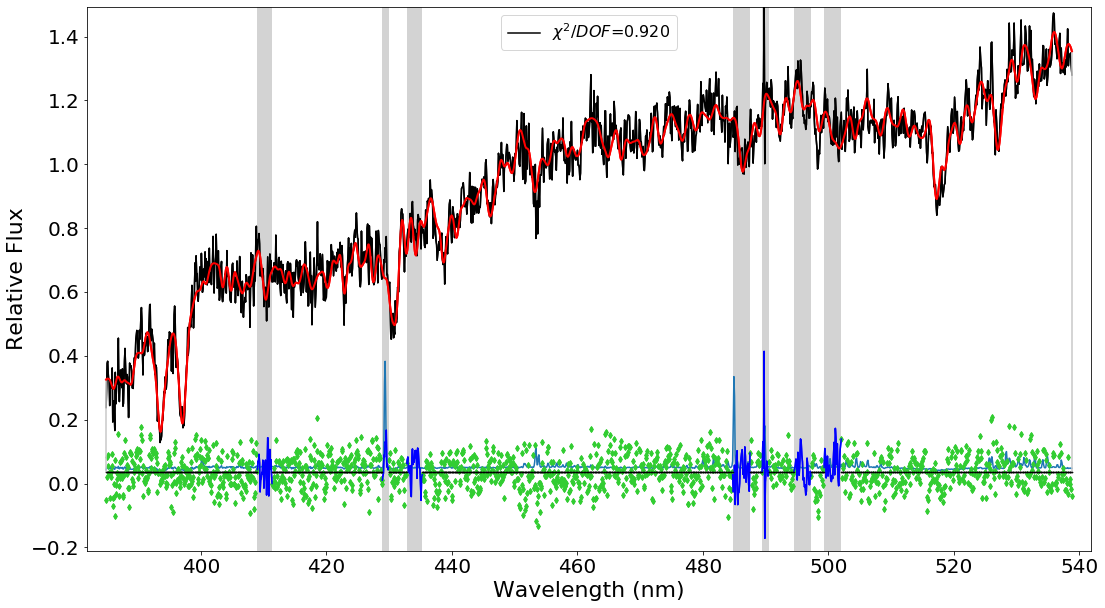}
    \caption{Central bin spectrum in the observer rest-frame. Observed data are in black, model obtained using \texttt{pPXF} is shown in red, grey vertical lines are masked regions, and the residual is shown green. The best fit parameters are $V = 141 \pm 9$\,km/s and $\sigma_V = 246 \pm 10$\,km/s, with the uncertainties given by the bootstrapping approach. In the top centre, the reduced $\chi^2$. For details of the spectral fitting, see the text.}
    \label{fig:bin0}
\end{figure}

We fit the two first moments of the Gauss-Hermite polynomials $[V, \sigma_V]$, including an additive polynomial of order 4 for each Voronoi bin. Moreover, we mask possible emissions lines and two regions where the variance in the data was very prominent. We perform our modelling in the observer rest-frame, which means that we bring the galaxy spectra to the rest-frame before the fitting. The result for the central bin can be seen in Figure~\ref{fig:bin0}.

To measure the uncertainty in the fitted parameters, we use a bootstrapping approach, and the strategy is the following. A first fit is performed, called best fit, from where the kinematic measurements are obtained, and the residuals for each pixel are computed. Now, from each pixel in the best fit spectrum, we use the value as a centre of a Gaussian distribution and the respective residual as the dispersion of this distribution. This ensures a new value for each pixel in the spectrum, drawn from the Gaussian distribution. Following this procedure for all the pixels, a new full spectrum is generated. Then we fit this new spectrum with \texttt{pPXF} (setting the bias flag equal to zero) and save the outputs $[V, \sigma_V]$. We repeat this procedure generating 200 new spectra and measuring $V$ and $\sigma_V$. After that, we compute the $1\sigma$ dispersion of each parameter, and use it as the uncertainty associated with the measurement of that parameter\footnote{This strategy is very similar to other bootstrapping approaches used by the community. In particular, our approach is based on two main codes. One by Jonathan Cohn (\url{https://github.com/jhcohn/ppxf/blob/master/ppxf_nifs_kinematics_witherr.py}), and the other by Remington Oliver Sexton (\url{https://github.com/remingtonsexton/BADASS3/blob/master/badass3_v7_7_6.py}).}.

After fitting all the Voronoi bins, we construct the $V_\text{rms} = \sqrt{V^2 + \sigma_V^2}$ map, which is the root-mean-square velocity for each Voronoi bin. The central velocity, i.e., the velocity of the central bin, was subtracted from each of the Voronoi bin velocities to obtain the real velocity with respect to the centre of the galaxy. We do not see any evidence of rotation. This $V_\text{rms}$ map is shown in Figure~\ref{fig:muse-data}.

We use the following error propagation to estimate the uncertainties in the $V_\text{rms}$
\begin{equation}
    1\sigma_{\text{rms}} = \frac{\sqrt{ (V \times 1\sigma_{\text{vel}})^2 + (\sigma_V \times 1\sigma_{\text{disp}})^2   }}{V_\text{rms}},
\end{equation}
where $1\sigma_{\text{rms}}$ is the  uncertainty in the root-mean-square velocity, $1\sigma_{\text{vel}}$ is the $1\sigma$ uncertainty in the velocity, and $1\sigma_{\text{disp}}$ the $1\sigma$ uncertainty in the velocity dispersion.

\subsection{ALMA data}\label{sub:Alma-data}
High-resolution interferometric data of SDP.81  were obtained as part of Science Verification for the 2014 ALMA Long Baseline Campaign, using the most extended configuration of the instrument to that date. These observations were taken in bands 4 ($\sim 2$\,mm), 6 ($\sim 1.3$\,mm) and 7 ($\sim 1.0$\,mm) using  between 23 and 36 antennas, and are publicly available in the ALMA Science Portal\footnote{\url{https://almascience.eso.org/}} (ASP). These bands comprise both continuum emission and some emission lines of CO and H$_2$O. The data reduction and calibration were performed following the pipeline provided by ASP and implemented using the  Common Astronomy Software Applications \citep[\texttt{CASA};][]{2007ASPC..376..127M}. More information about observation, data reduction and imaging can be found in \citet{Vlahakis_2015}.

In this work, we perform the lens modelling directly in the image plane, instead of the $uv$-plane (see Section~\ref{Sec. Modelling}). We use the image of continuum band 7, which includes the $\sim 250\mu$m rest-frame emission, for the lens modelling. The image was also binned from a pixel scale of $0.005\arcsec $ to a pixel scale of $0.01\arcsec $, to increase the modelling efficiency and reduce the covariance between the pixels \citep[][]{Dye2015}. During the modelling, we assumed the synthesised beam sizes described in the ALMA data: $0.0308\arcsec  \times 0.0235\arcsec $ ($\sim 141\text{\,pc} \times 108\text{\,pc}$ at the lens redshift), position angle (PA) of $15^{\degr}$, counterclockwise from North. Moreover, the error map was estimated by the rms background value, measured using \texttt{CASA} in a region where no emission is seen, with the addition of $10$\% of the value of each pixel summed in quadrature to account for possible calibration errors. 

\begin{figure}
	\includegraphics[width=\columnwidth]{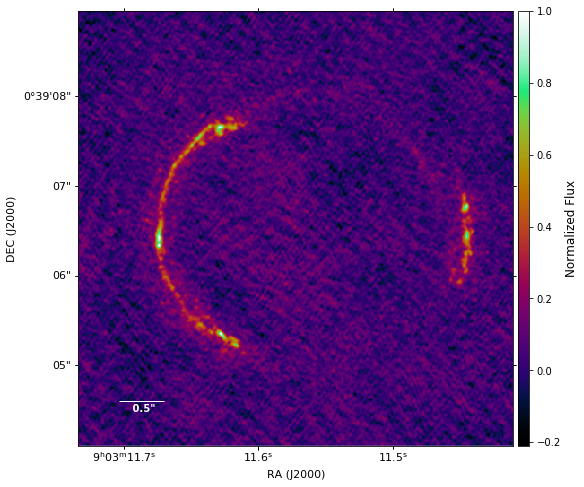}
    \caption{ALMA observations of SDP.81 system, where only the background emission of the source is visible, due to the spectral window of the observations (band 7 continuum). The typical flux emission is in the order of mJy. North is up, and East is left. }
    \label{fig:alma}
\end{figure}

Figure~\ref{fig:alma} shows the emission of the lensed source in ALMA band 7. Due to the wavelength observed by ALMA, the lens light is not visible, and only the deflected source is captured, making this dataset very suitable for lens modelling since the contamination due to the lens galaxy light is negligible. Panel (c) in Figure~\ref{fig:hst-data} show the near-infrared emission of the lensed source at $z=3.04$ observed with HST/F160W and ALMA band 7 emission contours overlaid. As it is immediately seen in the figure, there is a distinct offset between the submm (white contours, tracing the dust) and the near-infrared emission (mainly tracing the stars). An additional structure is observed to the North of the HST image and a much larger extent of the F160W emission of the western gravitational arc towards the South. Such offset and extended near-infrared emission were also reported by \citet{Dye2015}, where the authors conclude, after the source reconstruction, that the most plausible scenario to explain these observations is that the background source consists of two objects which are merging. The authors also argue that this galaxy merger scenario is supported by the reconstructed source kinematics (from the CO emission), which reveals a rotating disc of gas and dust in a state of collapse.

\section{Galaxy Mass Profile}\label{Sec. Mass Profile}
To probe GR as intended in this paper, we need to combine lensing and kinematics information, and for that, a self-consistent mass model is required. Assuming a total mass profile for both models we can simultaneously fit it to the lensing and kinematic data.

The galaxy mass profile employed for the analysis of the SDP.81 lens galaxy can be divided into: (i) a stellar mass component, obtained by deprojecting the observed surface brightness profile; (ii) a dark matter halo component, described by a  Navarro-Frank-White \citep[NFW;][]{Navarro_1997} profile, which is motivated by N-body simulations.

Hereafter we employ the following notation for the right-handed coordinate systems: $(R, \phi, z)$ are cylindrical coordinates;  $(x^{\prime}, y^{\prime}, z^{\prime})$ are projected coordinates assuming $z^{\prime}-$axis along the line of sight, $(x^{\prime}, y^{\prime})$ in the plane of the sky, and $x^{\prime}-$axis along the galaxy projected semi-major axis; the origin of both systems is the galaxy centre, which assumes that the galaxy has an axisymmetric geometry. The galaxy inclination $i$ is the angle between $z-$ and $z^{\prime}-$axes, such that  $i = 90^{\circ}$ implies that the galaxy is edge-on. 

\subsection{Stellar mass profile}
To account for the contribution of the stellar content in the mass profile, we use the Multi-Gaussian Expansion \citep[MGE;][]{Emsellem1994, Cappellari2002} parametrization, which consists in modelling the observed projected surface brightness profile of a galaxy as a sum of two-dimensional elliptical concentric Gaussians. Assuming that the mass follows the light, the parametrized surface brightness, in turn, can be used as a tracer for the stellar mass density profile. One advantage of this method is its capability to capture deviations of the isophotes from ellipses, making possible the modelling of multi-component objects such as lenticulars. Furthermore, the intrinsic stellar luminosity density (i.e. the deprojected quantity) has an analytical form for this parametrization. At the same time, the gravitational potential can be evaluated using a single and simple Gaussian quadrature.

Assuming that $I(x^{\prime},y^{\prime})$ is the projected surface brightness, we can parametrize it as follows

\begin{equation}\label{eq:surf_MGE}
    I(x^{\prime},y^{\prime}) = \sum_{j=1}^{N} \frac{L_j}{2\pi \sigma_{j}^{2} q^{\prime}_j}\exp{\left[-\frac{1}{2\pi \sigma_{j}^{2}} \left(x_{j}^{\prime 2} + \frac{y_{j}^{\prime 2}}{q_{j}^{\prime 2}}\right)\right]},
\end{equation}
where $N$ is the total number of Gaussians adopted. The $j^{\text{th}}$ Gaussian component has a total luminosity $L_j$, an observed projected axial ratio $0 \leq q^{\prime}_j  \leq 1$, and a dispersion $\sigma_j$ along the semi-major axis, which is aligned with $x^{\prime}$-coordinate.

To obtain the intrinsic three-dimensional luminosity density $\nu$ it is necessary to deproject the luminosity surface density given by equation~(\ref{eq:surf_MGE}). However,  the deprojection is not unique, except in the cases where the galaxy is edge-on \citep[e.g.][]{1988MNRAS.231..285F}. Fortunately, the deprojection becomes unique once a model is adopted. Following \citet{Cappellari2002}, and assuming an oblate axisymmetric model, the intrinsic luminosity density can be written, in cylindrical coordinates, as

\begin{equation}\label{eq:Intrinsic Lumunisosity Density}
    \nu(R, z) = \sum_{j=1}^N \frac{L_j}{(2\pi)^{3/2} \sigma_{j}^3 q_j} \exp{\left[-\frac{1}{2\pi \sigma_j^2} \left(R^2 + \frac{z^2}{q_{j}^2}\right)\right]},   
\end{equation}
where $L_j$ and $\sigma_j$ are the same as in (\ref{eq:surf_MGE}), and $q_j$ is the deprojected three-dimensional intrinsic axial ratio, related to the projected axial ratio by
\begin{equation}\label{eq: q deproj}
    q_{j}^2 = \frac{q_{j}^{\prime 2} - \cos^2{i}}{\sin^2{i}}.
\end{equation}

The luminosity density can be easily converted to mass density, assuming a $M/L$. Although many works show a good agreement between data and model when a constant $M/L$ is assumed \citep[e.g.][]{Williams_2009,Atlas3d_2013}, for a robust approach, we can assume one $M/L$ for each Gaussian component

\begin{equation}
    \Upsilon_j = M_j/L_j,
\end{equation}
where $M_j$ is the mass of the $j^{\text{th}}$ Gaussian component with $L_j$ luminosity. Therefore, the density mass profile is given by

\begin{equation}\label{eq:MGE mass density}
    \rho(R, z) = \sum_{j=1}^N \frac{M_j}{(2\pi)^{3/2} \sigma_{j}^3 q_j} \exp{\left[-\frac{1}{2\pi \sigma_j^2} \left(R^2 + \frac{z^2}{q_{j}^2}\right)\right]}.  
\end{equation}

Once we have obtained the mass density profile, the gravitational potential can be derived from the Homoeoid Theorem for densities stratified on similar concentric ellipsoids \citep[][]{Chandrasekhar1969, Binney_Tremaine2008} as

\begin{equation}\label{eq: Gravitational Potential}
    \Phi(R,z) = -G\sqrt{\frac{2}{\pi}}\sum_{j=1}^{N}\frac{M_j}{\sigma_j}\Tilde{\Phi}_j(R,z),
\end{equation}
where $G$ is the gravitational constant and $\Tilde{\Phi}_j(R,z)$ is given by

\begin{equation}
    \Tilde{\Phi}_j(R,z) = \int_{0}^{1} \frac{d\tau}{\sqrt{1 - \zeta_{j}^{2} \tau^{2}}} \exp{\left[-\frac{\tau^2}{2\sigma_j^2} \left(R^2 + \frac{z^2}{1 - \zeta_{j}^{2} \tau^{2}}\right)\right]},
\end{equation}
with $\zeta_{j}^{2} = 1- q_{j}^{2}$.

Instead of using a different $M/L$ for each MGE component, which can introduce a lot of free parameters in our model, we opt to use a $M/L$ modulated by a Gaussian function\footnote{To the best of our knowledge, such approach was first implemented in \url{https://github.com/HongyuLi2016/JAM}.
}, which allows us to set a different $M/L$ for each component of the MGE parametrization based on its dispersion, however with a small number of free parameters. In addition, this modulation ensures that the total baryonic $M/L$ profile is monotonically decreasing, consistent with observations of real galaxies.
The modulated-Gaussian $M/L$ has three parameters: the central stellar mass-to-light ratio $\Upsilon_0$, a gradient parameter $\delta$ that describes the smoothness of the Gaussian, and a parameter $\upsilon_0$, which is the ratio between the central and the outermost $M/L$. This can be expressed as

\begin{equation}\label{eq: Gaussian ML}
    \Upsilon_j = \Upsilon_0 \left[\upsilon_0 + (1 - \upsilon_0)e^{ -0.5 (\sigma_j \, \delta)^2}\right],
\end{equation}
where $\sigma_j$ is the dispersion associated with the MGE component. 

Finally, using the mass density distribution (equation~\ref{eq:MGE mass density}), and the associated potential (equation~\ref{eq: Gravitational Potential}), we can solve the necessary equations (see Section~\ref{Sec. Modelling}) for the lensing and dynamical modelling using relatively simple expressions.

\subsection{Dark matter halo}
To represent a possible dark matter halo, we include a NFW mass density profile 

\begin{equation}\label{eq: NFW mass profile}
    \rho(r) = \frac{\rho_s}{(r/r_s)(1 +  r/r_s )^2} = \frac{\kappa_s \Sigma_{\text{{\tiny crit}}}}{r(1 + r/r_s)^2 },
\end{equation}
where $\rho_s$ is the characteristic density, $r_s$ is the scale radius, and $\kappa_s = \frac{\rho_s r_s}{\Sigma_{\text{{\tiny crit}}}}$. The critical surface density $\Sigma_{\text{{\tiny crit}}}$ is given by 
\begin{equation}\label{eq:critical_density}
	\Sigma_{\text{{\tiny crit}}} \equiv \frac{c^2}{4\pi G}\frac{D_S}{D_L D_{LS}},
\end{equation}
where $D_{LS}$, $D_L$ and $D_S$ are the angular diameter  distances between the lens and source, lens and observer, and source and observer, respectively, in the lensing context (see Section~\ref{Sec. Modelling}).

Typically, this profile has two free parameters: the scale radius $r_s$, and the overall amplitude $\kappa_s$, where both can be a function of the mass range, cosmology and redshift \citep{Wyithe_2001, Diemer_2013}.

We can still introduce an ellipticity in this mass profile, replacing the radial coordinate $r$ by a elliptical version $ r_q = \sqrt{R^2 + z^2/q^2_{\text{\tiny{DM}}}}$, where  $q_{\text{\tiny{DM}}}$ is the three-dimensional axial ratio\footnote{For $q_{\text{\tiny{DM}}} < 1$ the halo is oblate, for $q_{\text{\tiny{DM}}} = 1$ spherical and for $q_{\text{\tiny{DM}}} > 1$ prolate}. In this paper we assume that the halo is aligned with the stellar mass profile, so the orientation of the halo is fixed. Then, the dark matter mass profile has three free parameters, the scale radius $r_s$, the amplitude $\kappa_s$,  and the axial ratio $q_{\text{\tiny{DM}}}$ of the halo.

To easily include the dark matter profile in both models (lens and dynamics), it is convenient to parametrize the elliptical dark matter halo using the MGE approach \citep[e.g.][]{Atlas3d_2013, Li2016}. Making use of this approach the mass profile and the gravitational potential are still given by the equations~(\ref{eq:MGE mass density}) and~(\ref{eq: Gravitational Potential}), however with the sum performed over $N + N_{\text{\tiny{DM}}}$, where $N_{\text{\tiny{DM}}}$ is the number of Gaussians used to parametrize the dark matter halo.

\section{Modelling}\label{Sec. Modelling}
This section describes the modelling procedures and how we combine both datasets in a self-consistent model. We also explore the Bayesian inference employed for the sampling the parameter space.

\subsection{Lens modelling}\label{lens-modelling}

Traditionally, lens modelling can be performed using two different approaches, considering a parametric or a pixelated source surface brightness. In the former, given a parametric light distribution and a mass model for the lens galaxy, each pixel in the source-plane can be traced to the image-plane through the lens equation \citep[see, e.g.][]{Schneider1992, Meneghetti2016}, and the result compared with the observed lensed source \citep[e.g.][]{LENSED2016}. While in the latter, given a mass model for the lens galaxy and the lensed source, we are able, in principle, to reconstruct the source object by ray tracing back, pixel by pixel, the observed surface brightness of the lensed image, inverting the lens equation and assuming a pixelated source plane. This second approach is known as the semi-linear inversion method \citep[SLI;][]{SLI2003}. In both cases, the PSF effects can be taken into account.

When the source light profile is more complex, parametric surface brightness profiles may not reproduce the real source morphology. To overcome this issue, the pixelated source plane and the SLI can be employed. Furthermore, as shown by \citet{Nightingale2015}, adaptive gridding, such as Voronoi pixelization, has some advantages over regular gridding, such as removing some significant biases, albeit at the cost of more computational time.

Once both approaches are important to our analysis (Section~\ref{Sec:Pipeline}), we employ the \texttt{PyAutoLens} \citep{Nightingale2021} package, which is a suite for the SGL modelling developed in Python. For details of its implementation, modelling and source reconstruction, see \citet{Nightingale2018}.

Throughout our modelling, the goodness of fit depends on the model adopted: when a parametric source is considered, the goodness of fit is given by a simple $\chi^2$ statistics, and when a pixelated source is considered, the goodness of fit is placed within a Bayesian framework  \citep[][]{Suyu2006} and computed through the lens model evidence $\epsilon$.

The key parameter in the lens modelling is the deflection angle $\bmath{\alpha} = (\alpha_{x^\prime},\alpha_{y^\prime})$, which is associated to the mass density distribution of the lens galaxy  \citep[see e.g.][]{Schneider1992}, and the geometry of the system, i.e., the three angular diameter distances $D_{LS}$, $D_L$ and $D_S$. In the case of the MGE mass profile, the deflection angle can be described as \citep{VanDeVen2010,Barnabe2012}

\begin{multline}\label{deflection_x}
    \alpha_{x^{\prime}}(x^{\prime}, y^{\prime}) = \frac{1}{\pi D_L^2 \Sigma_{\text{{\tiny crit}}}} \int_0^1 \tau d\tau \sum_{k} \frac{M_k}{\sigma_k} \frac{\Tilde{x}^{\prime}}{\sqrt{1 - \eta_k^2\tau^2}} \times \\ 
     \times\exp\left[ -\frac{\tau^2}{2} \left( \Tilde{x}^{\prime 2} + \frac{\Tilde{y}^{\prime 2}}{1 - \eta_k^2\tau^2}\right)\right], 
\end{multline}

\begin{multline}\label{deflection_y}
    \alpha_{y^{\prime}}(x^{\prime}, y^{\prime}) = \frac{1}{\pi D_L^2 \Sigma_{\text{{\tiny crit}}}} \int_0^1 \tau d\tau \sum_{k} \frac{M_k}{\sigma_k} \frac{\Tilde{y}^{\prime}}{(1 - \eta_k^2\tau^2)^{3/2}} \times \\ 
     \times\exp\left[ -\frac{\tau^2}{2} \left( \Tilde{x}^{\prime 2} + \frac{\Tilde{y}^{\prime 2}}{1 - \eta_k^2\tau^2}\right)\right], 
\end{multline}
where we define $\Tilde{x}^{\prime} \equiv x^{\prime}/\sigma_k$, $\Tilde{y}^{\prime} \equiv y^{\prime}/\sigma_k$ and $\eta_k^2 = (1 - q_k^{\prime 2})$. In these expressions, the deflection angle and $\sigma_k$ are given in radians, and $\Sigma_{\text{{\tiny crit}}}$ is the critical density given by equation~(\ref{eq:critical_density}). The summation is over the number of Gaussian that represents the mass density profile.

Unfortunately, \texttt{PyAutoLens} does not provide the MGE parametrization as an alternative to light or mass profile. For this reason, we implement the MGE deflection angle in the context of the \texttt{PyAutoLens} framework, whilst the light and dark matter mass parametrizations were performed using the \texttt{MgeFit} package \citep{Cappellari2002}.

The deflection angle described by the equations (\ref{deflection_x}) and \ref{deflection_y}) assumes GR, and therefore, assumes $\eta_{\text{PPN}} = 1$. However, considering that the gravitational slip parameter can probe deviations from GR, the observed deflection angle is related to the deflection angle predicted by GR as

\begin{equation}
    \bmath{\alpha} = \frac{1 + \eta_{\text{PPN}}}{2}\bmath{\alpha}_\text{GR},
\end{equation}
where $\bmath{\alpha}$ is the observed deflection angle, and $\bmath{\alpha}_\text{GR}$ is the deflection angle predicted by GR. Since the deflection angle is related to the mass profile, we can use it, in association with equation~(\ref{eq: PPN masses}), to probe GR using lensing and dynamical measurements (\citetalias{Collett2018}).

In this work, we opt to perform the modelling in the image plane instead of the visibility plane of ALMA data since it allows us to mask and fit only the regions where the emission of the lensed source is detected. Furthermore, this choice makes the modelling more computationally efficient (we model thousands of pixels instead of millions of visibilities) and improves the goodness of fit since extended areas of background sky are left aside. The main disadvantage of this choice is that we are modelling the Fourier transform of the data instead of the interferometric visibilities. As a consequence of the procedure of obtaining an image from the visibilities, the image pixels become correlated by the beam, which could bias the image-plane modelling if the uncertainties do not take into account the covariance between the pixels. However, due to the high coverage of the $uv-$plane in these ALMA data, the error associated with the image plane is significantly reduced. In addition, \citet{Dye_2018} performs a comparison between modelling the cleaned image data and the visibility data directly, where they found only minor differences in the reconstructed source and lens model when an SLI is applied. They also find that these differences become more notable when the coverage of the $uv-$plane becomes more sparse. Nevertheless, it is worth mentioning that the data used in that work has a much lower angular resolution than the data obtained for the SDP.81 system, mainly due to ALMA configuration and integration time.

\subsection{Kinematic modelling}
The dynamical state of a collisionless system (e.g. an elliptical galaxy without interactions) can be described by the Collisionless Boltzmann equation \citep[][]{Binney_Tremaine2008}. However, the distribution function (DF) that characterises this system is a function of seven parameters (three positions, three velocities and time). The typical available data are not sufficient to recover this seven-dimensional DF. As a consequence, some simplifications are required to describe the dynamical state of the system. Considering a steady-state axisymmetric configuration, the system should satisfy the two Jeans equations in cylindrical coordinates \citep[][]{Binney_Tremaine2008,Cappellari2008}:

\begin{equation}\label{eq: Jeans1}
    \frac{\partial (\nu \overline{v^2_R})}{\partial R} 
    + 
    \frac{\partial (\nu \overline{v_R \, v_z})}{\partial z}
    +
    \nu \left[  \frac{\overline{v^2_R} - \overline{v^2_\phi}}{R} + \frac{\partial \Phi}{\partial R}     \right]
    =
    0,
\end{equation}

\begin{equation}\label{eq: Jeans2}
   \frac{1}{R}\frac{\partial \left(R \nu \overline{v_R \, v_z}\right)}{\partial R} 
    + 
    \frac{\partial (\nu \overline{v^2_z})}{\partial z}
    +
    \nu \frac{\partial \Phi}{\partial z}
    =
    0,
\end{equation}
where

\begin{equation}
    \nu \overline{v_i \, v_j} = \int{v_i v_j f d^3 \bmath{v}},
\end{equation}
and $f$ is the DF of the stars, $\Phi$ is the total gravitational potential, $(v_R, v_\phi, v_z)$ are the velocities for each cylindrical coordinates,  and  $\nu$ is the intrinsic luminosity density.

As previously stated, our mass model can take into account contributions from stellar and dark matter components, and both contributions can be represented by a sum of two-dimensional elliptical concentric Gaussians (Section~\ref{Sec. Mass Profile}). In this case, the potential is given by equation~(\ref{eq: Gravitational Potential}), and the luminosity distribution by equation~(\ref{eq:Intrinsic Lumunisosity Density}).

Even for a fixed $\Phi$ and $\nu$, the two Jeans equations~(\ref{eq: Jeans1}) and~(\ref{eq: Jeans2}), still depend on four unknown quantities ($\overline{v^2_R}$, $\overline{v^2_\phi}$, $\overline{v^2_z}$, and $\overline{v_R \, v_z}$), such that additional assumptions are required  in order to determine a unique solution. Therefore, we assume the same prescription made in \citet{Cappellari2008}: (i) the velocity dispersion ellipsoid is aligned with the cylindrical coordinate system (all the off diagonal terms vanish); and (ii) a constant flattening of the orbits in the meridional plane, i.e. $\overline{v^2_R} = b \overline{v^2_z}$, where $b$ is the anisotropy. Often the anisotropy parameter in the $z$ direction, $\beta_z$, is rewritten as

\begin{equation}
    \beta_z = 1 - {\overline{v^2_z}}\, / \, {\overline{v^2_R}} \equiv 1 - \frac{1}{b}.
\end{equation}
This particular parametrization is convenient when the MGE approach is employed, once each Gaussian component could have its anisotropy $\beta_i$ \citep[see Eq. (24) of ][]{Cappellari2008}. However, individual $\beta_i$ are not physical quantities \textit{per se}.

With such assumptions, the Jeans Equations reduce to 
\begin{equation}\label{eq: Final_Jeans1}
    \frac{\partial (b\nu\overline{v^2_z})}{\partial R} 
    + 
    \nu \left[  \frac{b\overline{v^2_z} - \overline{v^2_\phi}}{R} + \frac{\partial \Phi}{\partial R}     \right]
    =
    0,
\end{equation}

\begin{equation}\label{eq: Final_Jeans2}
    \frac{\partial (\nu \overline{v^2_z})}{\partial z}
    +
    \nu \frac{\partial \Phi}{\partial z}
    =
    0.
\end{equation}
Imposing the boundary condition $\nu\overline{v^2_z} = 0$  when $z \rightarrow \infty$, the solution can be written as 
\begin{equation}\label{eq:Jeans v_phi}
    \overline{v^2_\phi} = b \left[ \frac{R}{\nu} \, \frac{\partial (\nu \, \overline{v^2_z})}{\partial R} + \overline{v^2_z} \right] + R\frac{\partial \Phi}{dR}
\end{equation}
\begin{equation}\label{eq:Jeans v_z}
    \overline{v^2_z}  = \frac{1}{\nu}\, {\displaystyle \int_{z}}^\infty d\zeta \, \nu \frac{\partial \Phi}{d\zeta},
\end{equation}
where $\zeta$ is the integration variable.

Finally, these intrinsic quantities should be integrated along the line-of-sight in order to obtain the projected second velocity moment $\overline{v^2_{\text{los}}}$, which can be directly compared with the stellar kinematic observables, i.e. the root-mean-square velocity $V_{\text{rms}}$ (Section~\ref{Sec. Data}). The goodness of fit, in turn, can be computed as a $\chi^2$ statistic between data and model.

To solve the axisymmetric Jeans equations and compute the projected second velocity moment, we use the Jeans Anisotropic Modelling \citep[\texttt{JAM};][]{Cappellari2008, Cappellari2020}, in its Python implementation. Since \texttt{JAM} is designed to operate within the MGE framework, the gravitational potential $\Phi$ is described by the same sum of elliptical Gaussians as in the lens modelling, equation~(\ref{eq: Gravitational Potential}).

\subsection{Pipeline prescription}\label{Sec:Pipeline}
A well-known issue within the SLI method is under/over-magnified solutions for the lens mass model. When one of these solutions is achieved, the reconstructed source resembles an under/over-magnified version of the observed source image instead of the unlensed compact source. Although these solutions present poor goodness of fit, they occupy a large volume in the parameter space and, in some cases, can even correspond to strong evidence $\epsilon$ \citep[see Fig. 3 of ][]{Maresca2021}, which can be a problem for most of the sampling algorithms.

In order to suppress these erroneous solutions, multiple techniques have been employed, such as careful tuning priors, auto-identification using machine learning algorithms \citep{Maresca2021}, and pipelines that first fit a parametric profile to the source galaxy and then use these results to shrink the parameter space and avoid unwanted solutions during the SLI process \citep{Nightingale2018}. In this paper, we employ the latter method.

Our pipeline is composed of five phases, each focusing on different aspects of the modelling. In the following, we describe each phase of the pipeline.

\textbullet \, \textbf{Phase 1} ({Ph1}): \textit{Parametric Source, Lens + Dynamical modelling} - To avoid the under/over-magnified solutions, we initialize the modelling considering a parametric source profile, for which these unphysical solutions do not exist. The main idea is to estimate the region in the parameter space where the mass profile does not generate under/over-magnified solutions. Therefore, in this phase, we fit the MGE lens mass profile + \texttt{JAM} model + Sérsic source profile. The Sérsic profile has seven parameters: source centre ($x^\prime_0, y^\prime_0$), axial ratio ($q_{\text{source}}$), orientation angle ($\phi_{\text{source}}$) defined counterclockwise from the positive $x^\prime$-axis, the intensity ($I_{\text{source}}$), the effective radius ($R_\text{eff}$),  and the Sérsic index ($n$). The parameters for the lens model and the dynamical model may vary depending on the model adopted, i.e., a constant $M/L$ or a Gaussian $M/L$, one or multiple anisotropies etc. The $\eta_{\text{PPN}}$ parameter is also sampled during this phase as part of the lens model.  We assume non-informative priors for all the parameters, except for the intensity (a log-uniform prior), source centre (Gaussian priors centred at the origin) and the $\eta_{\text{PPN}}$ parameter (Gaussian prior centred at $1.00$ and dispersion equal to $0.09$). We choose to assume a Gaussian prior over $\eta_{\text{PPN}}$, with dispersion consistent with the uncertainty present in \citetalias{Collett2018}, since both works are focused in extragalactic tests of GR and the expected (if it happens) small deviations from GR between the systems at this scale.

The sampling results in the most probable values ({MP}$_1$), which are defined by the median of the one-dimensional marginalised posterior probability distribution (see Bayesian inference in Section~\ref{sub:Bayesian}).

\textbullet \, \textbf{Phase 2} ({Ph2}): \textit{Adaptive Pixelization and Hyperparameters} - In this phase we fix the parameters of the MGE lens mass profile and \texttt{JAM} model in the most probable values from {Ph1}, {MP}$_1$. We use an adaptive Voronoi grid to reconstruct the background object. The grid of pixels is represented by an irregular grid of Voronoi cells, allowing any shape, size or tesselation. We also apply a constant regularisation term responsible for penalizing sources solutions that are less smooth \citep[see e.g.][]{SLI2003,dye2008}. Therefore, we left the hyperparameters regularisation ($\lambda$) and source grid shape ($x,y$) as free parameters with non-informative priors. The main goal of this phase is to initialize the inversion process.

\textbullet \, \textbf{Phase 3} ({Ph3}): \textit{Model Refinement} I - In this phase, the MGE lens mass model and the \texttt{JAM} model are refined. To that, we fix the source inversion hyperparameters of  {Ph2} in the most probable values ({MP}$_2$) and leave the lens and dynamical parameters to vary. During this phase, we update the prior knowledge of the parameters such that the new priors have the same form (non-informative or Gaussian) as the previous one. However, the new non-informative priors are defined by {MP}$_1 \pm 20\%$  or  {MP}$_1 \pm 1\sigma$, whichever defines a larger interval. The Gaussian prior is updated such that the new centre is equal to the previous MP result, and the dispersion equal to $10\%$ of the {MP}$_1$ or $1\sigma$, whichever is greater. This approach reduces the parameter space and alleviates the under/over-magnified solutions. Also, this phase is important to capture the true morphology of the source (through the SLI).

\textbullet \, \textbf{Phase 4} ({Ph4}): \textit{Adaptive Brightness-based Pixelization and Hyperparameters} - In this phase, we employ an adaptive Brightness-based Voronoi Pixelization grid with a constant regularisation term, once again in order to reconstruct the source. Unlike from {Ph2}, in this step, the pixelization adapts to the source surface brightness, therefore reconstructing areas of high flux with higher resolution\footnote{Justifying the needed of {Ph3}, in order to capture the true morphology of the source galaxy.}. This Brightness-based Pixelization is related to a weighted K-means clustering algorithm, which has three hyperparameters: the total number of pixels in the Voronoi pixelization, i.e., the number of K-means clusters ($N_\text{pix}$), and two parameters that control the weighting, the weight floor ($W_f$), and the weight power ($W_p$). For a precise definition of these parameters and methods, see the discussion present in Sec. 4.7 of \citet{Nightingale2018}. Besides these three parameters, we resample the regularisation hyperparameter $\lambda$ since we apply a new discretization method. During the sampling of these new hyperparameters, we keep fixed the lens and dynamical parameters in the {MP}$_3$ result.

\textbullet \, \textbf{Phase 5} ({Ph5}): \textit{Model Refinement} II - In this phase, we search for the final inference of the parameters, fixing the {MP}$_4$ result, and allowing the lens and dynamical parameters to vary. However, recently \citet{Etherington2022} demonstrated that the discretization during the SGL modelling using \texttt{PyAutoLens} is subject of bias. This bias is mainly determined by the random seed that determines the centres of the Voronoi source pixels during the K-means clustering and is translated into ``spikes'' in the figure of merit of the model (see their Fig. 3). These spikes could confuse the sampler, which can be stuck in tiny volumes of the multi-dimensional parameter space and thus underestimate the total statistical uncertainty. To overcome this issue, \citeauthor{Etherington2022} propose a likelihood cap value, from which the lens likelihood cannot be greater. This cap value is determined by a bootstrapping approach. Using the {MP}$_3$ and {MP}$_4$ results, we compute 300 likelihood evaluations, each of them using a different K-means seed. This process produces a distribution, which can be fitted using a Gaussian profile, whose mean will act as the likelihood cap. It is important to note that this cap is only for the lens model since the bias is related to the discretization of the source plane. Using this cap, we are finally able to refine our model. We update the priors in the same way as in Ph3. Non-informative priors are defined by  {MP}$_3 \pm 10\%$  or  {MP}$_3 \pm 1\sigma$, whichever defines a larger interval and the Gaussian prior is centred in  {MP}$_3$ result with dispersion equal to $10\%$ of {MP}$_3$ or its $1\sigma$, whichever is greater. Since the goal during this phase is to improve the most probable parameters and estimate their statistical uncertainty, we chose to define a small non-informative prior using only $10\%$ of the values of the parameters, as opposed to the $20\%$ used before (although with a conservative dispersion for the $\eta_{\text{PPN}}$). Nevertheless, since only a small subset of the parameters provides plausible physical solutions for the lens model, the parameters typically have small statistical uncertainty, being dominated by systematic effects (see  Section~\ref{Sec. Discussion}). Furthermore, during this phase, if the lens likelihood computed is above the cap defined above, it is pushed down to the cap's value before it is returned to the sampler. This, of course, does not prevent the likelihood associated with the dynamic model from growing such that the joint likelihood can assume values above the cap. Finally, the most probable model is given by the median of the one-dimensional marginalized posterior probability distribution of the {Ph5} sampling, {MP}$_5$.

A table summarising the parameters and their priors is presented in Appendix~\ref{Ap: Priors}.

\subsection{Bayesian inference}\label{sub:Bayesian}
In order to estimate the distribution of the parameters and infer their posterior probability, we employ a Bayesian inference, which is a well-known way to estimate the posterior probability $P(\bmath{\Theta}|\bmath{D},\bmath{M})$ of a set of parameters $\bmath{\Theta}$ for a given model $\bmath{M}$ conditioned on some data $\bmath{D}$. This information can be accessed through the Bayes' rule

\begin{equation}\label{eq: Bayes Rule}
    P(\bmath{\Theta}|\bmath{D},\bmath{M}) = \frac{P(\bmath{D}|\bmath{\Theta},\bmath{M}) P(\bmath{\Theta}|\bmath{M})}{P(\bmath{D}|\bmath{M})} \equiv  \frac{ \bmath{\mathcal{L}}(\bmath{\Theta_M}) \bmath{\pi}(\bmath{\Theta_M}) }{\bmath{\mathcal{Z}_M}},
\end{equation}
where $P(\bmath{D}|\bmath{\Theta},\bmath{M}) \equiv \bmath{\mathcal{L}}(\bmath{\Theta_M})$ is the likelihood of the model given the data, $P(\bmath{\Theta}|\bmath{M}) \equiv \bmath{\pi}(\bmath{\Theta_M})$ is the prior of the parameters which quantifies our initial knowledge about them, and 

\begin{equation}
    \bmath{\mathcal{Z}_M} = \int_{\bmath{V_{\Theta}}}  P(\bmath{D}|\bmath{\Theta},\bmath{M})P(\bmath{\Theta}|\bmath{M})d\bmath{\Theta}
\end{equation}
is the marginal likelihood for the data, given a model. The integral is over the entire parameter space $\bmath{V_{\Theta}}$, i.e.,  over all possible parameter combinations. 

We can obtain the most probable set of parameters that reproduce the data through the posterior distribution of the parameters, given a model and the priors. We define the most probable values of the parameters as the median value of the one-dimensional marginalised posterior probability distribution of the individual parameters and quantify the statistical uncertainty as to their $68\%$ credible intervals (by taking the 16th and 84th percentiles), which are roughly equivalent to the $1\sigma$ uncertainty.

To explore the parameter space and sample the posterior, we use the public, open-source \texttt{dynesty} \citep{Speagle_2020}, a Python implementation of the Nested Sampling algorithm  \citep{Skilling2006}. \texttt{dynesty} offers an extensive range of customisations for the sampling procedure. Once our modelling is divided into five phases, we define different combinations for each pipeline phase to improve computational efficiency.

For all pipeline phases, we use the static version of \texttt{dynesty}. We employ a multiple ellipsoids bounding distribution \citep{Mukherjee_2006, Feroz_2019}, which can better handle with multi-modal distributions, and a random walk proposal sampling \citep[][]{Metropolis,Hastings1970} with $N_{\text{walks}} = 15$. Since in the first two phases, we are only interested in estimating a marginal region of the parameter space where the model has physical solutions, we define the termination criteria \citep[see, e.g.][]{Skilling2006} as \texttt{dlogz}$= 10$. On the other hand, in the last three phases, the termination criteria are \texttt{dlogz}$= 0.8$. Moreover, as the multi-dimensional parameter space has many dimensions, we use a total of 180 live points in each phase, except for {Ph2}, which runs with 100 live points.

\subsection{Combined modelling}
Since the lensing and the kinematical data are independent, we can consider the individual likelihoods of each model separately, such that we can produce a single likelihood for the combined modelling by multiplying the respective likelihoods:

\begin{equation}
    \bmath{\mathcal{L}}_{\text{Model}} \equiv \bmath{\mathcal{L}}_{\text{Lens}}\times \bmath{\mathcal{L}}_{\text{Dyn}},
\end{equation}
where $\bmath{\mathcal{L}}_{\text{Lens}}$ is the likelihood of the lens model, and $\bmath{\mathcal{L}}_{\text{Dyn}}$ is the likelihood of the dynamical model. During {Ph1} the lens likelihood is quantified by a $\chi^2$ statistic \citep[e.g.][]{SLI2003}, while in {Ph2}-{Ph5} by the evidence $\epsilon$ \citep[e.g.][]{Suyu2006, Nightingale2015}. The likelihood of the dynamical model, in turn, is giving by a $\chi^2$ statistic between the observed $V_{\text{rms}}$ map and the \texttt{JAM} model \citep[e.g.][]{Li2016}.

To account for the effects of PSF, we parametrize both HST and MUSE point spread functions as a sum of 2D-circular Gaussians, such that they can be used for the light decomposition of the SDP.81 lens galaxy and during the dynamical modelling.

\texttt{JAM} model and the deflection angle, equations~(\ref{deflection_x}) and (\ref{deflection_y}), assumes that the $x^\prime$-axis is coincident with the galaxy projected major axis. To fulfil this requirement, we rotate the kinematical and the ALMA data by the position angle of $11.9\degr$, determined by the \texttt{FIND\_GALAXY} algorithm  \citep[][]{Cappellari2002}. We also assume that the dark matter centre,  the luminous mass centre, and the lens light centre coincide.

We assume that the stellar mass component follows the observed luminosity distribution of the lens galaxy, modulated by a Gaussian $M/L$ given by equation~(\ref{eq: Gaussian ML}). The MGE parametrization describes the lens luminosity distribution (Section~\ref{Sec. Mass Profile}) of the HST/F160W image, which is used for lens and dynamical modelling. The MGE fit results in 9 Gaussian components described in Table~\ref{table:MGE}. Therefore, the stellar component has 3 free parameters ($\Upsilon_0, \delta, \upsilon_0$).

The dark matter content is described by an elliptical NFW profile, equation~(\ref{eq: NFW mass profile}), which is parametrize as a sum of elliptical Gaussians (MGE approach), such that both lens and dynamical models can take it into account. However, for computational efficiency, instead of including the Gaussians describing the dark matter content in the deflection angle (equations~\ref{deflection_x} and \ref{deflection_y}), we take advantage of the fact that the deflection angle is linear with respect to mass, such that the contribution coming from different mass profiles can be added directly. Thus, the inclusion of the dark matter component in the lens models uses the NFW deflection angle directly \citep[e.g][]{Keeton2001, Nightingale2018}, instead of its MGE parametrization. 

We carefully analysed the error associated with the MGE parametrization of the elliptical NFW profile and how it impacts the deflection angle. We conclude that the error related to this parametrization in the deflection angle is of the order of $0.04\%$. Such a small error shows that using the elliptical NFW profile instead of its MGE parametrized version does not significantly impact the final result.

Since we do not have enough data to constrain the dark matter halo on larger scales, we fix $r_s$ as 10 times the effective radius\footnote{We measure it using the approach described by \citet{Atlas3d_2013}} ($R_{\text{eff}} =  1.14\arcsec$) of the stellar component \citep[][]{Kravtsov2013,Sonnenfeld2015}. We also assume that the halo has the same orientation as the stellar component, thus the dark matter component has only two free parameters $\kappa_s$ and $q_{\text{\tiny{DM}}}$. 

Additionally, we do not include the contribution of a supermassive black hole (SMBH), since the spatial resolution of the IFU data would not allow imposing strong constrains on this parameter. However, we test a model with the inclusion of a SMBH, aiming to verify the impact of this parameter on systematic uncertainties associated with the mass model (see Section \ref{Sec. Discussion}). 
The model has other five parameters: two associated to a possible external shear in the lens model, the shear magnitude ($\text{shear}_{\text{mag}}$), and the shear orientation ($\text{shear}_{\phi}$), measured counterclockwise from the positive $x^\prime$-axis; the galaxy inclination $i$, restricted by the equation~(\ref{eq: q deproj}); the orbital anisotropy $\beta_z$, which is considered constant ($\beta_i = \beta_z$); and the $\eta_{\text{PPN}}$ parameter itself.

\begin{table}
\centering
\caption{MGE components of the HST/F110W image of SDP.81 lens galaxy. We convert the MGE units of counts into physical units as described by \citet{Trick2016}. We use the Vega zero-point and the absolute solar magnitude from \citet{Willmer_2018}. The columns are, in order, the galaxy projected surface density, the Gaussian dispersion, and the observed axial ratio for each MGE component.
}
\label{table:MGE}
\begin{tabular}{lcc} 
\hline
 $I$ [L$_\odot$/pc$^2$]  & $\sigma$ [arcsec]& $q^\prime$\\
\hline
17964.30	&	   0.05	 &  0.72	\\ 
5062.47		&	   0.13	 &  0.62	\\
728.03		&	   0.26	 &  0.82	\\ 
277.24		&	   0.55	 &  1.00	\\ 
139.13		&	   0.69	 &  0.47	\\ 
91.66		&	   1.07	 &  1.00	\\ 
40.06		&	   1.48	 &  0.57	\\ 
29.55		&	   2.65	 &  0.61	\\ 
17.58		&	   2.65	 &  1.00    \\ [1ex]
\hline
\end{tabular}
\end{table}

Besides the parameters described above, additional parameters are sampled together during some phases. For instance,  the parameters describing the parametric source ({Ph1}), and the hyperparameters of Ph2 and Ph4. However, since we have little interest in these parameters, we will not discuss them in this paper.

To summarise, our fiducial model during the {Ph5}  has ten free parameters: three related to the stellar mass component ($\Upsilon_0, \delta, \upsilon_0$); two related to the dark matter halo component ($\kappa_s, q_{\text{\tiny{DM}}}$); one inclination ($i$); one parameter describing the anisotropy ($\beta_z$); two describing the effects of a possible external shear in the lens model ($\text{shear}_{\text{mag}}, \text{shear}_{\phi}$); and the gravitational slip parameter ($\eta_{\text{PPN}}$).
\section{Main Results}\label{Sec. Results}
\begin{figure}
	\includegraphics[width=\columnwidth]{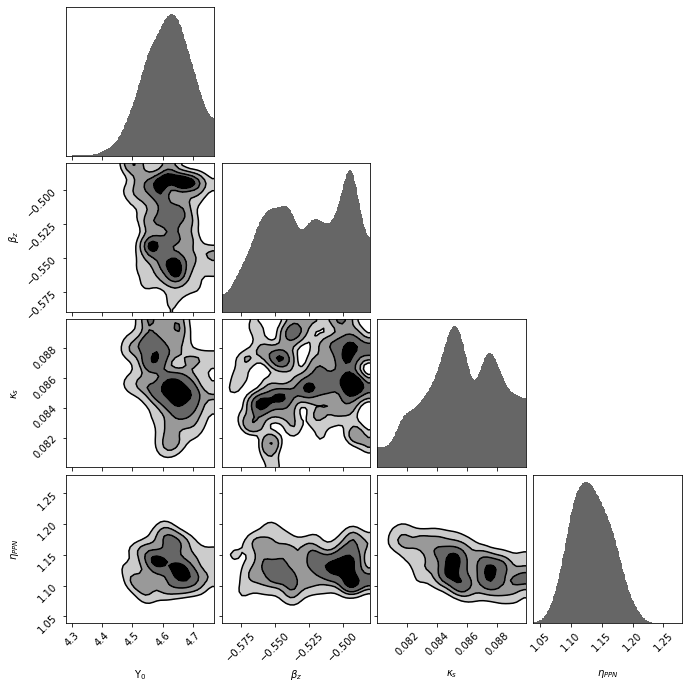}
    \caption{Two-dimensional posterior distributions for the parameters of the fiducial model. The complete list with the most probable values and their associated statistical uncertainty can be found in  Table~\ref{table:Ph5 results}. The contours roughly correspond to  $0.5\sigma$, $1\sigma$, $1.5\sigma$, and $2\sigma$.}
    \label{fig:fidual_posterior}
\end{figure}

The fiducial model was obtained by fitting simultaneously both the observed lensing and kinematical data. The {MP}$_5$ values and their associated statistical uncertainties are summarised in Table~\ref{table:Ph5 results}, while the two-dimensional posterior distributions for some of the parameters are shown in Figure~\ref{fig:fidual_posterior}.

\begin{table}
\centering
\caption{Most probable values and $68\%$ confidence intervals for the parameters of the fiducial model. For the description of the parameters see Sec.~\ref{Sec. Modelling}. All angles are measured with respect to the reference frame rotated by the PA $11.9\degr$. }
\label{table:Ph5 results}
\begin{tabular}{lcc} 
\hline
 Parameter  & \textbf{MP}$_5$ & Physical Units\\
\hline 
$\Upsilon_0$	&   $4.63^{+0.06}_{-0.09}$	  & M$_\odot$/L$_\odot$  \\  &&\\
$\delta$		&	$1.78^{+0.09}_{-0.09}$	  & arcsec$^{-1}$        \\  &&\\
$\upsilon_0$	&	$0.88^{+0.07}_{-0.05}$	  & -         \\  &&\\
$i$		        &	$82^{+4}_{-4}$   & degree               \\  &&\\ 
$\beta_z$		&	$-0.53^{+0.03}_{-0.04}$   & -                   \\  &&\\
$\kappa_s$		&	$0.086^{+0.002}_{-0.003}$ & -                   \\  &&\\
$q_\text{DM}$	&	$0.49^{+0.02}_{-0.01}$	  & -                   \\  &&\\ 
$\eta_{\text{PPN}}$&$1.13^{+0.03}_{-0.03}$	  & -                   \\  &&\\ 
$\text{shear}_{\text{mag}}$&$0.022^{+0.001}_{-0.001}$ & -           \\  &&\\
$\text{shear}_{\phi}$&$55^{+2}_{-3}$& degree                \\ [1ex]
\hline
\end{tabular}
\end{table}

\begin{figure*}
    \centering
    \subfloat{
    \includegraphics[width=0.75\columnwidth]{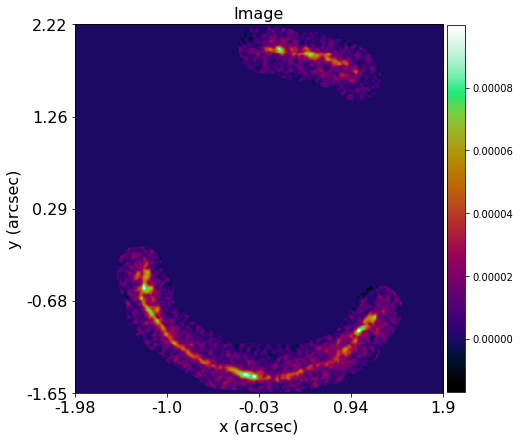}
    }
    \quad 
    \subfloat{
    \includegraphics[width=0.75\columnwidth]{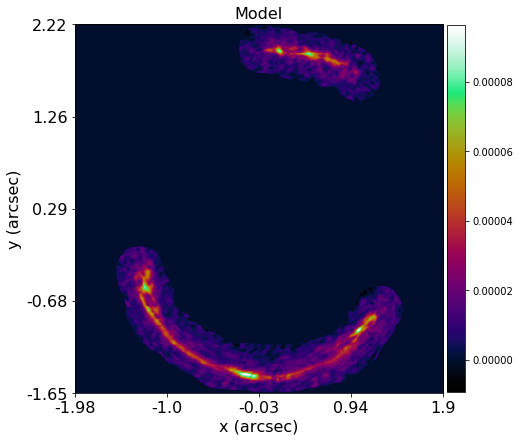}
    }
    \\
    \subfloat{
    \includegraphics[width=0.75\columnwidth]{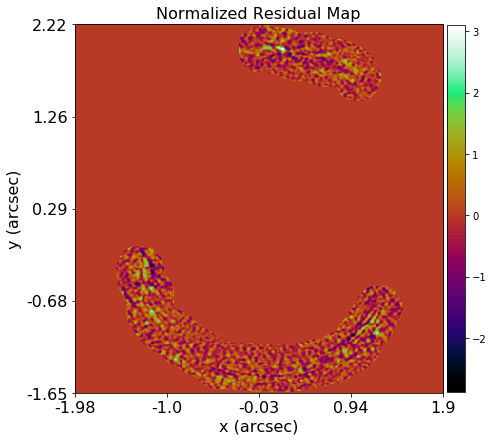}
    }
    \quad
    \subfloat{
    \includegraphics[width=0.75\columnwidth]{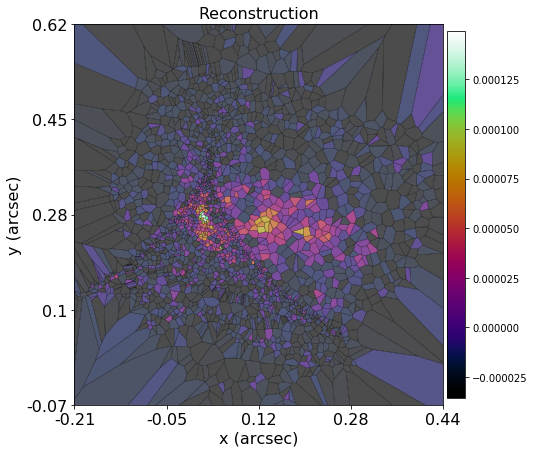}
    }
   \caption{The most probable lens model assuming the fiducial model. From the top left to the bottom right panel: the ALMA image data, the lensed source reconstruction, the normalised residual map (data - model)/noise, and the reconstructed source galaxy. Note that all the panels are rotated by the position angle, as required by \texttt{JAM}  and the deflection angle. The colour bars are in units of mJy.}
    \label{fig:fidual_lens_model}
\end{figure*}

\begin{figure*}
	\includegraphics[width=\textwidth]{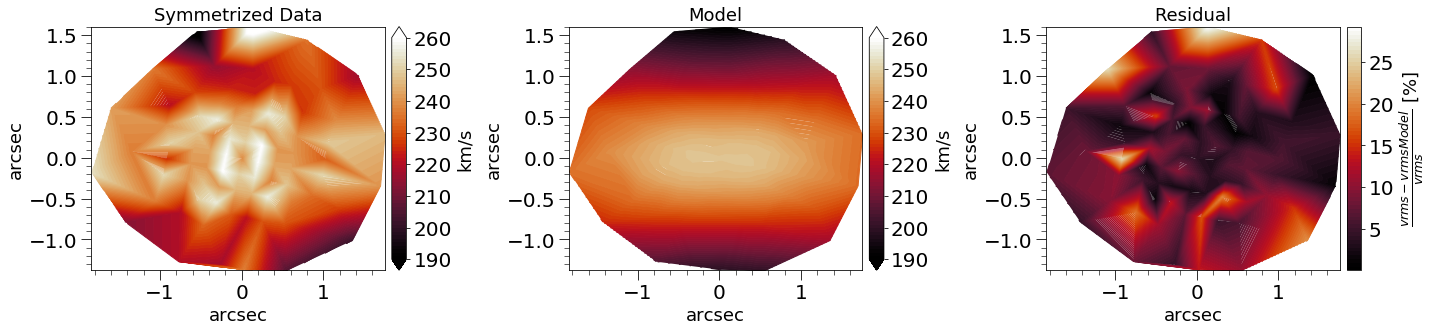}
    \caption{Most probable kinematical model assuming the fiducial configuration. From left to right: the symmetrized $V_\text{rms}$ map using only the Voronoi bins, the $V_\text{rms}$ map obtained from the most probable dynamical model, and the residuals. All figures are linearly interpolated for better visualisation, and rotated by the position angle as required by \texttt{JAM}  and the deflection angle.}
    \label{fig:fidual_dyn_model}
\end{figure*}

Due to the complexity of the model adopted we can derive different properties of the lens galaxy, and at the same time, reconstruct the source emission. The most probable lens model and the reconstructed source are shown in Figure~\ref{fig:fidual_lens_model}, and the most probable dynamical model in Figure~\ref{fig:fidual_dyn_model}. 

We estimate the Einstein radius ($R_{\text{Ein}}$) using the \texttt{PyAutoLens} routine \texttt{einstein\_radius\_from}, which defines the $R_{\text{Ein}}$ as the radius of a circle with the same area as the area within the tangential critical curve. We found $R_{\text{Ein}}  = 1.61\arcsec$, which agrees with previous studies \citep[e.g.,][]{Dye2014, Vlahakis_2015,Wong2015}.

We have found a relatively higher ($4.63$~M$_\odot$/L$_\odot$) $M/L$ (on average within $R_{\text{Ein}}$) when compared with previous works of SDP.81 \citep[e.g][]{Wong2015, Tamura2015}. However, we can address this discrepancy to the fact that our $M/L$ is modulated by a Gaussian function, allowing the $M/L$ to vary over different scales in the galaxy, as opposed to the previous works, which considered only a constant value. Furthermore, as we will show in the next section, even when assuming a constant $M/L$, our composite model seems to prefer higher values than those previously found.

We notice a moderate degeneracy between the central $M/L$, $\Upsilon_0$, and the $\eta_{\text{PPN}}$, as shown in Figure~\ref{fig:fidual_posterior}. However, a more prominent degeneracy between $\kappa_s$ and these other two parameters is seen. A stronger degeneracy appears in the relation between the $\beta_z$ and $\Upsilon_0$. Nevertheless, as evidenced by the anisotropy posterior distribution, this parameter is not well constrained by the kinematical data.

Assuming the fiducial parameters, we use the \texttt{MGETOOLS}\footnote{\url{https://github.com/lauralwatkins/mgetools}} package to estimate the fraction of dark matter inside $R_{\text{Ein}}$. We have found that the dark matter fraction, $f_{\text{DM}}$, is of the order of  $\sim 35\%$, showing that the galaxy is baryonic dominated in the inner regions. Our result is consistent with results for other strong lenses at similar redshift \citep[e.g.][]{Auger_2010, Sonnenfeld2015}. However, we have found a smaller fraction than \citet{Wong2015}, which assumes a spherical NFW and infers a dark matter fraction of the order of $\sim 50\%$ for the SDP.81 galaxy within $R_{\text{Ein}}$.

As expected by the lens data quality, the shear components are well constrained and in agreement with the previous works \citep[e.g.][]{Dye2015, Wong2015, Tamura2015}. 

The dynamical model, in turn, reproduce the features of the data, as shown by the third panel in Figure~\ref{fig:fidual_dyn_model}. On the other hand, the fractional error increases through the outskirts, which is expected since the SNR rapidly decreases with the galactic radius. At the same time, this imposes severe limitations in order to constrain the dark matter scale radius, justifying our choice of keeping it fixed in our fiducial model.

With the current spectroscopic data, we do not expect to improve the constraints on the dynamics or even in the $M/L$, however with better IFS data, stellar population models could be used to narrow the priors in both cases, attempting to capture any possible gradient in the $M/L$ \citep[e.g][]{Poci2022}.

Finally, our fiducial inference to $\eta_{\text{PPN}}$ is excluding GR beyond $1\sigma$ confidence level. Meanwhile, as demonstrated by previous works with a similar methodology, systematic uncertainties are the main source of uncertainty on $\eta_{\text{PPN}}$. In the next section, we discuss this problem.
\section{Discussion and Systematics}\label{Sec. Discussion}
As shown by previous works \citep[e.g.][]{Pizzuti2016, Cao2017, Collett2018}, the main source of uncertainty in this kind of analysis is systematic uncertainties. The systematics could be related to different factors, for instance, the mass model adopted, the cosmology assumed, and our lack of knowledge about the kinematic properties of the lens galaxy.

To account for these uncertainties, we apply the same methodology described  in Section~\ref{Sec. Modelling}, for alternative mass models, and we stress our fiducial model for the upper and lower limits of the adopted cosmology, i.e., we verify how the uncertainty on the Hubble constant impacts the inference of $\eta_{\text{PPN}}$.  We also verify how the measurement of the stellar velocity dispersion is affected by the choice of a particular stellar library.

All the {Ph5} results for each of the models discussed below are in Appendix~\ref{Ap: alternative} with their statistical $1\sigma$ uncertainty.

\subsection{Alternative Mass Profiles}
\begin{figure}
	\includegraphics[width=\columnwidth]{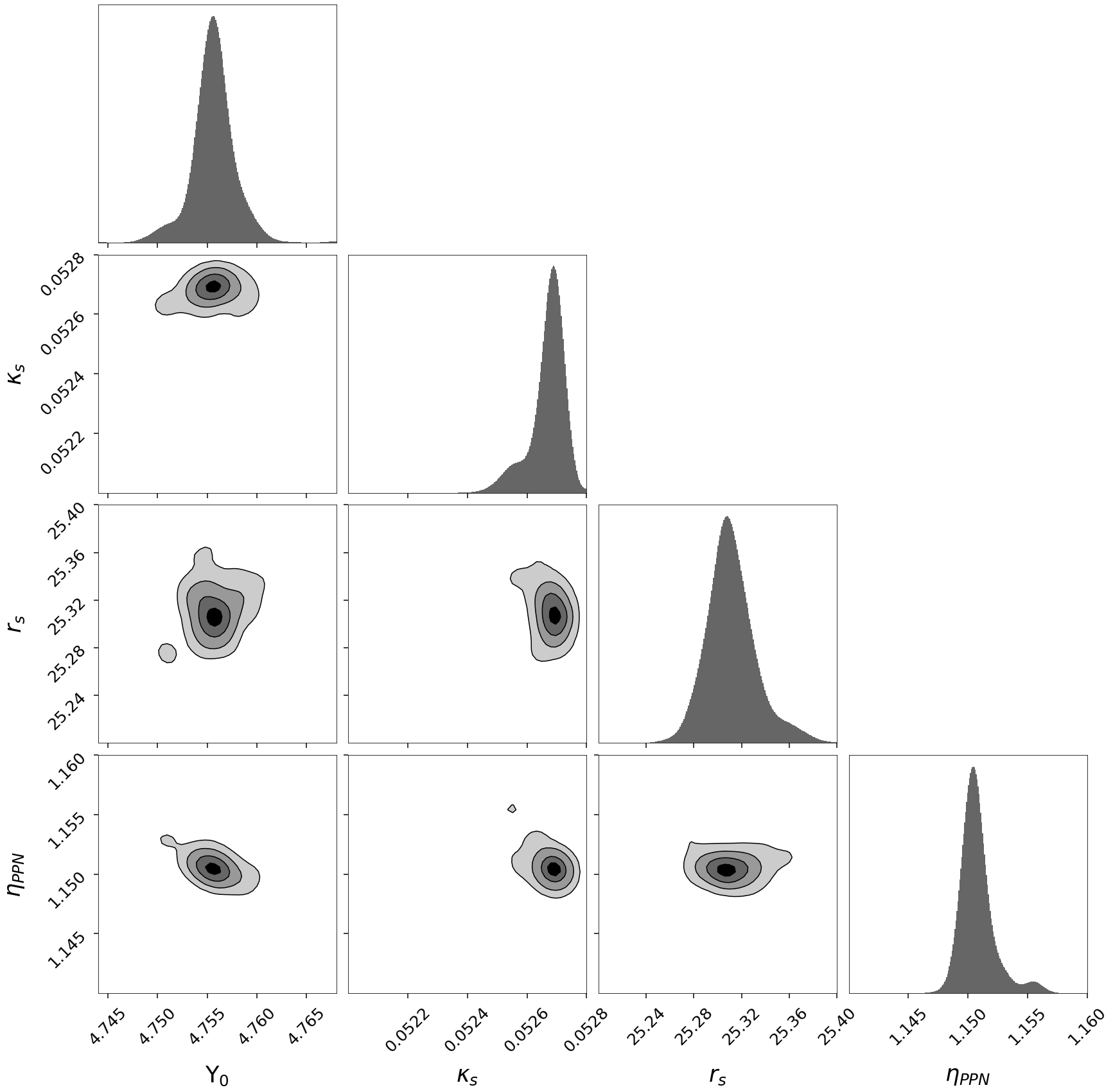}
    \caption{Two-dimensional posterior distribution for the parameters of the model with free dark matter scale radius. The contours roughly correspond to  $0.5\sigma$, $1\sigma$, $1.5\sigma$, and $2\sigma$.}
    \label{fig:pd_model20}
\end{figure}

The fact that we fix the dark matter scale radius in our fiducial model could be a source of systematics in our inference of  $\eta_{\text{PPN}}$. To quantify how this could impact the inference on $\eta_{\text{PPN}}$, we performed two more modelling using the fiducial configuration. First, we fix the dark matter radius to be 5 times the effective radius, for which we infer a $\eta_{\text{PPN}} = 1.12^{+0.03}_{-0.02}$ , and a fraction of dark matter within the Einstein ring consistent with $f_{\text{DM}} \sim 35\%$. Next, although our data do not cover the possible extended halo of dark matter, we consider the dark matter scale radius as a free parameter. This models gives $\eta_{\text{PPN}} = 1.151^{+0.008}_{-0.004}$, with a dark matter scale radius $r_s = 25.30^{+0.03}_{-0.02}$ arcsec, and a dark matter fraction inside the Einstein ring of the order of $\sim 28\%$. Figure~\ref{fig:pd_model20} shows the posterior distribution for some of the parameters of the model with a free dark matter radius.

To test different composite models, we perform the modelling assuming a similar configuration as \citet{Wong2015}, which assume a spherical dark matter halo, a constant $M/L$, and the contribution of a SMBH at the centre of the lens galaxy. For this  model, we apply the same priors, during {Ph1}, as \citet{Wong2015} to the parameters $r_s$, $\text{shear}_{\text{mag}}$, and the mass of the SMBH. For the further phases, we updated these priors as follows: for {Ph3}, the flat priors are updated such that the new flat priors are {MP}$_1\pm 10\%$ or {MP}$_1\pm 1\sigma$, whichever defines a greater interval; and for the  Gaussian priors the new mean is defined by the {MP}$_1$ result and the dispersion is equal to $10\%$ of the mean or its $1\sigma$, whichever is greater. For {Ph5} we follow this same procedure, but replacing the {MP}$_1$ by {MP}$_3$. All the other phases follow the same methodology described in Section~\ref{Sec. Modelling}. In our model, we include the SMBH as described by \citet{Cappellari2008}, representing the SMBH as a small circular Gaussian at the centre of the galaxy\footnote{In practice this means adding an extra Gaussian component to our MGE potential, equation~(\ref{eq: Gravitational Potential}).}.

For the mass model similar to \citet[][]{Wong2015}, we found that the dark matter halo radius consistent with $r_s = 19.19^{+0.05}_{-0.08}$ arcsec. This leads to a small fraction of dark matter inside the Einstein ring ($f_{\text{DM}} \sim 18\%$) when compared with the fiducial result. The mass of the SMBH ($\log_{10}{\frac{M_{\text{\tiny{SMBH}}}}{M_\odot}} = 6.80^{+0.02}_{-0.01}$) is significantly smaller than the one reported by \citet[][]{Wong2015} by almost 2 dex. Using this model we infer $\eta_{\text{PPN}} = 1.11^{+0.01}_{-0.02}$.

Taking the standard deviation between the different values of $\eta_{\text{PPN}}$ for these alternative mass models, we can estimate an additional systematic uncertainty of the order of $0.017$ on the slip parameter.

\subsection{Cosmology}
We derive our fiducial model under the assumption of $\Lambda$CDM and the most probable parameters of Planck mission \citep{Planck_2015}. However, the uncertainty on the Hubble constant could affect the angular diameter distances, which appear in both lens and dynamical modelling. In the dynamical modelling, the distance is used to obtain the intrinsic properties of the galaxy, while in the lens modelling, a combination of the three angular diameter distances appears in the critical surface density, equation~(\ref{eq:critical_density}). 

Since we are in a regime where the linearity of the Hubble law no longer holds, we are not able to scale the uncertainty on $H_0$ directly to $\eta_{\text{PPN}}$ as done by \citetalias{Collett2018}. Therefore, we opt to run the fiducial model again, however with different values of the Hubble constant, defined by the lower ($H_0 = 67.28$\,km\,s$^{-1}$\,Mpc$^{-1}$) and upper ($H_0 = 68.20$\,km\,s$^{-1}$\,Mpc$^{-1}$) values of the $1\sigma$ confidence level.

For the lower value of $H_0$, we infer $\eta_{\text{PPN}} = 1.05^{+0.04}_{-0.03}$. Using the upper value for $H_0$, we infer $\eta_{\text{PPN}} = 1.03^{+0.04}_{-0.04}$.

Contrary to the Planck results, which are based on measurements of the early Universe, some results using measurements from the late Universe, such as supernovae \citep[e.g.][]{Riess_2021} and time-delay cosmography \citep[e.g.][]{Millon_2020}, show a much higher value for the Hubble constant, in a statistical tension around\footnote{Or even above, depending on the dataset.} $4.5\sigma$ \citep[e.g.][]{Riess_2019,Di_Valentino_2021}. To see how (and if) this tension affects our result, we run three more models, assuming the fiducial configuration, however using values of $H_0$ consistent with those determined by time-delay cosmographic \citep[e.g.][]{Millon_2020, Shajib_2020}, namely $H_0 = 74.2 \pm 2$\,km\,s$^{-1}$\,Mpc$^{-1}$.

For the models assuming $H_0$ determined by  time-delay, we infer $\eta_{\text{PPN}} = (1.11^{+0.04}_{-0.04} \,; 1.13^{+0.02}_{-0.03} \,; 1.14^{+0.04}_{-0.04})$, for $H_0 = (74.2 \,; 72.2 \,; 76.2)$\,km\,s$^{-1}$\,Mpc$^{-1}$, respectively. All these results are consistent with the results derived using the Planck value of $H_0$, which is expected since our data are weakly sensitive to $H_0$, once the distance ratio effectively determines the Einstein radius, and $H_0$ effectively cancels.

Considering the standard deviation of the values of $\eta_{\text{PPN}}$ inferred for this six\footnote {5 derived using a different value of $H_0$ + fiducial model} values of $H_0$, we infer an additional uncertainty of the order of $0.043$ on the slip parameter, related to the uncertainty on the Hubble constant.

\subsection{Kinematics}
The lack of knowledge about the intrinsic kinematic properties of the lens galaxy is the main source of uncertainty in our analysis. Associated with this, our spectroscopy data do not have enough SNR to impose strong restrictions on the dynamic model. To estimate the level of uncertainty introduced by these issues, we evaluated how the inferred kinematics changes by the choice of a particular stellar template, re-deriving the kinematic map using the Medium resolution INT Library of Empirical Spectra \citep[MILES\footnote{\url{http://research.iac.es/proyecto/miles/}};][]{Miles_models} and the X-Shooter Spectral Library \citep[XSL\footnote{\url{http://xsl.u-strasbg.fr/index.html}};][]{XSL}.

For the XSL, we choose to use only the DR2 stars with UVB coverage, which matches our restframe spectra, and temperatures between $5000\,$K and $7000\,$K (mainly G and K stars), which roughly corresponds to the expected stellar population of an early-type galaxy. Since XSL has a spectral resolution higher than our restframe spectra, we can proceed as described in Section~\ref{Sec. Data}. On the other hand, the MILES library has a lower resolution than our restframe spectra, making the match of the resolution between the template and the instrument not possible before the fit. To deal with that, we perform the spectral modelling, and after that, we correct the measured $\sigma_\text{rms}$ by the quadratic differences in instrumental resolutions.

We found that the velocity dispersions measured by the XSL are systematically higher by a factor of $3.9\%$, on average than the inferred by the Indo-US templates. On the other hand, the inferred velocity dispersions using the MILES models are systematically smaller by $2.9\%$. However, all measurements agree within the uncertainties. 

Assuming the mean of these values, we expect a systematic of the order of $3.4\%$ due to poor kinematic constraints. This $3.4\%$ uncertainty on the velocity dispersion is responsible for a $13.6\%$ uncertainty on $\eta_{\text{PPN}}$.

\subsection{Final inference on $\eta_{\text{PPN}}$}
We now can combine in quadrature all the systematic uncertainties to estimate our final inference on the slip parameter: $0.017$ due to the effects related to the mass model adopted, $0.043$ related to the uncertainty on the cosmology, and $0.19$ uncertainty from the kinematic fit. We then infer that the most probable value for the slip parameter is $\eta_{\text{PPN}} = 1.13^{+0.03}_{-0.03}\pm0.20\,(\text{sys})$, which recovers GR within $1\sigma$ level.

The interesting here is to note that the posterior distributions for the slip parameter systematically prefer values above 1, can also be seen in previous, such as  in \citet{Liu_2022}, which uses a sample of intermediate redshift lens galaxies and their velocity dispersions. 

Unfortunately, due to systematic effects, we cannot assume any conclusion beyond this: that galaxies at high redshift appear to have a higher gravitational slip. To address and answer this issue, better spectroscopy data for intermediate systems are needed, once the systematic effects related to the kinematics are dominant in our inference, as in the previous analysis made by \citetalias{Collett2018}.

\section{Conclusions}\label{Sec. Conclusions}

Combining SGL and kinematics analysis has proven to be a straightforward way to probe GR, particularly on the galactic scales. In this paper, we consider the constraint on the $\eta_{\text{PPN}}$, defined by the ratio of two scalar potentials, using data from ALMA, MUSE and HST for the lens system SDP.81, with a lens galaxy at $z=0.299$.

We have performed this analysis assuming a self-consistent mass profile for the lens and dynamical modelling.  Our fiducial mass model considers the contribution of the stellar mass and the dark matter halo, represented by a NFW profile. Both contributions were parametrized by the MGE method, so that we were able to solve the Jeans equations and the lens equation assuming the same total mass profile. 

Our final inference is $\eta_{\text{PPN}} = 1.13^{+0.03}_{-0.03}\pm0.20\,(\text{sys})$, in accordance with GR predictions. In this final inference we take into account systematics related to the modelling, in order better understand the uncertainties. 

The uncertainty introduced by our lack of knowledge about the kinematic properties is of the order of $0.19$, and ally with the low SNR at the outskirts of the lens galaxy is the main source of systematics in our analysis.  The uncertainty in the Hubble constant seems to be a less problematic source of systematic, being responsible for only $0.043$. Although, assuming a cosmology derived under the $\Lambda$CDM can raise discussions about the recursion of testing GR using predictions of itself. An alternative to this problem is to derive the required distances through observations of type Ia (SN Ia) supernovae, as done by \citet{Liu_2022}. However,  it is necessary to have supernovae data covering the redshift range of both lens and source galaxies, which is currently limited by $z < 2.3$, considering the Pan-STARRS1 Medium Deep Survey sample \citep{Scolnic_2018}, which is far below the redshift of our source galaxy ($z_s = 3.042$).

In the future, to better understand the systematics, simulations can be used to quantify the sensitivity of the models. Hydrodynamical simulations, such as Illustris\footnote{\url{https://www.illustris-project.org/}}, can help to understand possible biases related to kinematic maps and maybe decide how many kinematic measurements are necessary for tight constraining the slip parameter at the per cent level. On the other hand, better IFS data will be available soon through the NirSpec instrument on board JWST, making possible more studies with intermediate galaxies and stronger kinematic constraints.

\section*{Acknowledgements}
The authors thank the referee for their comments and suggestions which led to an improved version of the manuscript. This project is funded by Conselho Nacional de Desenvolvimento Científico e Tecnológico (CNPq). The authors acknowledge the National Laboratory for Scientific Computing (LNCC/MCTI, Brazil) for providing HPC resources of the SDumont supercomputer, which have contributed to the research results reported within this paper. URL: \url{http://sdumont.lncc.br}.
This  work made use of the CHE cluster, managed and funded by COSMO/CBPF/MCTI, with financial  support  from  FINEP  and  FAPERJ,  and  operating  at  the  Javier  Magnin  Computing Center/CBPF. We thank Basílio Santiago , Davi Rodrigues, Rafael Nunes and Rogério Riffel for the fruitful discussion. We thank Thomas Wiecki (\texttt{CythonGSL}), Laura L. Watkins (\texttt{mgetools}), Hongyu Li, Jonathan Cohn, and Remington Oliver Sexton for making their codes publicly available. This paper makes use of the following ALMA data: ADS/JAO.ALMA \#2011.0.00016.SV. ALMA is a partnership of ESO (representing its member states), NSF (USA) and NINS (Japan), together with NRC (Canada), NSC, ASIAA (Taiwan) and KASI (Republic of Korea), in cooperation with the Republic of Chile. The Joint ALMA Observatory is operated by ESO, NAOJ and NRAO. This research has made use of the services of the ESO Science Archive Facility. Based on observations made with the NASA/ESA Hubble Space Telescope, and obtained from the Hubble Legacy Archive, which is a collaboration between the Space Telescope Science Institute (STScI/NASA), the Space Telescope European Coordinating Facility (ST-ECF/ESA) and the Canadian Astronomy Data Centre (CADC/NRC/CSA). CC acknowledges funding from CAPES and Conselho Nacional de Desenvolvimento
Científico e Tecnológico (CNPq). CF acknowledges funding from the CNPq through grants CNPq-314672/2020-6 and CNPq-433615/2018-4. ACS acknowledges funding from CNPq, the Rio Grande do Sul Research Foundation (FAPERGS) and the Chinese Academy of Sciences (CAS) President's International Fellowship Initiative (PIFI) through grants CNPq-11153/2018-6, CNPq-314301/2021-6, FAPERGS/CAPES 19/2551-0000696-9, 2021VMC0005.

\section*{SOFTWARE CITATIONS}
The scripts were written mostly for the Python Programming Language\footnote{\url{https://www.python.org/}}, using the Python version 3.7.6. They were also tested with Python version 3.6.3, and no bugs are reported. The  Operational system where the scripts were developed is Ubuntu 20.04.2 LTS (Focal Fossa)\footnote{\url{https://releases.ubuntu.com/20.04/}}, with architecture x86\_64.

This work uses the following software packages:

\begin{description}
\item \texttt{Astropy} \citep{astropy}
\item \texttt{CMasher} \citep{CMasher2020}
\item \texttt{dynesty} \citep{Speagle_2020}
\item \texttt{JAM}   \citep{Cappellari2008,Cappellari2020}
\item \texttt{Jupyter} \citep{jupyter}
\item \texttt{Matplotlib} \citep{Hunter:2007}
\item \texttt{MgeFit}  \citep{Cappellari2002}
\item \texttt{MPDAF}   \citep{mpdaf2017}
\item \texttt{Numba}   \citep{numba}
\item \texttt{NumPy}   \citep{numpy}
\item \texttt{pPXF}    \citep{Cappellari2016_PPXF}
\item \texttt{PyAutoLens} \citep{Nightingale2018,Nightingale2021}
\item \texttt{PyRAF}   \citep{Pyraf} 
\item \texttt{schwimmbad} \citep{schwimmbad} 
\item \texttt{SciPy}   \citep{Scipy_2020}
\item \texttt{VorBin} \citep{Vorbin_2003}
\end{description}

\section*{Data Availability}
The Hubble imaging data is publicly available at the Hubble Legacy Archive (\url{https://hla.stsci.edu/}) under the program-ID 12194 (PI: Negrello). The MUSE data is  available at ESO Science Archive Facility (\url{http://archive.eso.org/scienceportal/})  under the program-ID 294.B-5042 (PI: Gavazzi). The ALMA data is available through the ALMA Science Verification Program (\url{https://almascience.nao.ac.jp/alma-data/science-verification}).



\bibliographystyle{mnras}
\bibliography{example} 




\appendix

\section{Pipeline Priors}\label{Ap: Priors}

Table \ref{table: Parameters and Priors } describes the parameters and the priors applied during {Ph1} for the lens and dynamical modelling. The subsequent {Ph3} priors update are based on the results from {Ph1}, as well the {Ph5} priors are base on {Ph3} results. Typically, {Ph3} priors are {MP}$_1 \pm 20\%$  or  {MP}$_1 \pm 1\sigma$,  whichever defines a larger interval, while {Ph5} priors are {MP}$_3 \pm 10\%$ or {MP}$_3 \pm 1\sigma$, whichever defines a larger interval. Both $1\sigma$ are from the respective phase. Often, the $1\sigma$ interval is asymmetric. When it occurs, we choose the value that makes the prior broader.

Table \ref{table:Hyperparameters and Priors} describes the hyperparameters and their priors used during  {Ph2} and {Ph4}.

In the tables, $\mathcal{U}[a,b]$ denotes uniform (non-informative) priors between the lower value $a$ and the maximum value $b$;  $\mathcal{N}[a,b]$ denotes a normal (Gaussian) distribution with mean $a$ and dispersion $b$; $\mathcal{L}U[a,b]$ is a log-uniform distribution between the lower value $a$ and the maximum value $b$.

\begin{table}
\centering
 \caption{Parameters and priors used in this work. From left to right, the columns are: parameter symbol, prior applied during {Ph1}, parameter description, and physical unit. $^\textbf{a}$ the minimum value of inclination was determined using Eq.~(\ref{eq: q deproj}) and the lower value of the projected axial ratio (see Table \ref{table:MGE}). }
\label{table: Parameters and Priors }
 \begin{tabular}{lccc}
  \hline
 Parameter  & Prior & Description & Physical Unit\\
\hline
$x^\prime_0$           & $\mathcal{N}[0.0, 0.3]$   &   \makecell{Source $x^\prime$ centre} & arcsec\\  &&& \\
$y^\prime_0$           & $\mathcal{N}[0.0, 0.3]$   &   \makecell{Source $y^\prime$ centre} & arcsec\\  &&& \\
$q_{\text{source}}$ & $\mathcal{U}[0.1, 1.0]$   &   \makecell{Source axial ratio} & - \\  &&& \\
$\phi_{\text{source}}$ & $\mathcal{U}[0.0, 180.0]$   &   \makecell{Source orientation  \\ angle counterclockwise \\ from $x^{\prime}$-axis} & degree \\  &&& \\
$I_{\text{source}}$ & $\mathcal{L}U[10^{-6}, 10^{6}]$   &   \makecell{Source intensity} & e$^{-}$/s \\  &&& \\
$R_\text{eff}$& $\mathcal{U}[0.0, 30.0]$   &   \makecell{Effective radius} & arcsec \\  &&& \\
$n$& $\mathcal{U}[0.5, 8.0]$   &   \makecell{Sérsic index} & -\\  &&& \\
$\Upsilon$           & $\mathcal{U}[1.0, 15.0]$   &   \makecell{Constant $M/L$} & M$_\odot$/L$_\odot$\\  &&& \\
 $\Upsilon_0$           & $\mathcal{U}[1.0, 15.0]$   &   \makecell{Central $M/L$} & M$_\odot$/L$_\odot$ \\  &&& \\
 $\upsilon_0$           & $\mathcal{U}[0.0, 1.0]$   &   \makecell{Lower value \\ of $M/L$} & -\\  &&& \\
 $\delta$               & $\mathcal{U}[0.1, 2.0]$   &   \makecell{Smoothness of the \\ $M/L$ profile} & arcsec$^{-1}$\\ &&&  \\
 $\beta_z$            & $\mathcal{U}[-1.0, 0.5]$  &   Anisotropy & -\\  &&& \\
 $i$       & $\mathcal{U}[68.18, 90.0]^\textbf{a}$   &   \makecell{Galaxy \\ inclination}  & degree\\  &&& \\
 $\log_{10}{\frac{M_{\text{\tiny{SMBH}}}}{M_\odot}}$& $\mathcal{U}[6.5, 9.5]$  &   \makecell{Logarithmic of the mass  \\ of supermassive\\  black hole}  & -\\ &&&  \\
 $\kappa_s$             & $\mathcal{U}[0.0, 2.0]$   &   \makecell{Scale factor of \\ dark matter halo}  & -\\  &&& \\
 $r_s$                  & $\mathcal{U}[0.0, 50.0]$  &   \makecell{Scale radius of dark \\ matter halo} & arcsec\\  &&& \\
 $q_{\text{DM}}$        & $\mathcal{U}[0.4, 1.0]$   &   \makecell{Axial ratio of dark \\ matter halo} & -\\  &&&  \\
 $\text{shear}_{\text{mag}}$ & $\mathcal{U}[0.0, 0.1]$ &    Shear magnitude & -\\ &&& \\
 $\text{shear}_{\phi}$  & $\mathcal{U}[0.0, 180.0]$    &    \makecell{Shear angle \\ counterclockwise \\ from $x^\prime-$axis}& degree\\ &&& \\
 $\eta_{\text{PPN}}$  & $\mathcal{N}[1.00, 0.09]$ &    \makecell{Slip \\parameter} & -\\[1ex] 
 \hline
 \end{tabular}
\end{table}

\begin{table}
\centering
\caption{Hyperparameters and priors used in this work. From left to right, the columns are: parameter symbol, prior applied during {Ph2} and {Ph4}, and parameter description. } 
\label{table:Hyperparameters and Priors}
\begin{tabular}{lcc} 
\hline
 Hyperparameter  & Prior & Description \\
\hline
$\lambda$ & $\mathcal{L}U[10^{-6}, 10^{6}]$   &   \makecell{Constant \\ regularisation}\\
$x$ & $\mathcal{U}[20, 60]$   &   \makecell{$x$ shape of  the \\ source grid}\\
$y$ & $\mathcal{U}[20, 60]$   &   \makecell{$y$  shape of the \\ source grid}\\
$N_\text{pix}$ & $\mathcal{U}[50, 2500.00]$   &   \makecell{Number of K-means \\ clusters}\\
$W_f$ & $ \mathcal{U}[0.0, 1.0]$   &   \makecell{Weight floor }\\
$W_p$ & $ \mathcal{U}[0.0, 20.0]$   &  \makecell{Weight power }\\[1ex] 
\hline
\end{tabular}
\end{table}

\section{Results of the alternative models}\label{Ap: alternative}
Each of the following tables shows the result of  {Ph5} for the alternative models discussed in Sec.~\ref{Sec. Discussion}.

\begin{table}
\centering
\caption{Most probable values and $68\%$ confidence intervals for the  model with dark matter scale radius fixed as being $r_s = 5R_{\text{eff}}$. For the description of the parameters see Sec.~\ref{Sec. Modelling}.}
\begin{tabular}{lcc} 
\hline
 Parameter  & {MP}$_5$ & Physical Units\\
\hline 
$\Upsilon_0$	&   $4.34^{+0.08}_{-0.07}$	  & M$_\odot$/L$_\odot$  \\  &&\\
$\delta$		&	$0.20^{+0.01}_{-0.01}$	  & arcsec$^{-1}$        \\  &&\\
$\upsilon_0$	&	$0.87^{+0.05}_{-0.05}$	  & -             \\  &&\\
$i$		        &	$71^{+3}_{-3}$   & degree               \\  &&\\ 
$\beta_z$		&	$-0.46^{+0.01}_{-0.03}$   & -                   \\  &&\\
$\kappa_s$		&	$0.132^{+0.004}_{-0.006}$ & -                   \\  &&\\
$q_\text{DM}$	&	$0.45^{+0.03}_{-0.02}$	  & -                   \\  &&\\ 
$\eta_{\text{PPN}}$&$1.12^{+0.03}_{-0.02}$	& -                   \\  &&\\ 
$\text{shear}_{\text{mag}}$&$0.028^{+0.001}_{-0.002}$ & -           \\  &&\\
$\text{shear}_{\phi}$&$64^{+2}_{-2}$& degree                \\ [1ex]
\hline
\end{tabular}
\end{table}

\begin{table}
\centering
\caption{Most probable values and $68\%$ confidence intervals for the  model with free dark matter radius. For the description of the parameters see Sec.~\ref{Sec. Modelling}.}
\begin{tabular}{lcc} 
\hline
 Parameter  & {MP}$_5$ & Physical Units\\
\hline 
$\Upsilon_0$	&   $4.755^{+0.001}_{-0.002}$	  & M$_\odot$/L$_\odot$  \\  &&\\
$\delta$		&	$0.2071^{+0.0002}_{-0.0002}$	  & arcsec$^{-1}$        \\  &&\\
$\upsilon_0$	&	$0.789^{+0.001}_{-0.002}$	  & -             \\  &&\\
$i$		        &	$77.7^{+0.1}_{-0.1}$   & degree               \\  &&\\ 
$\beta_z$		&	$-0.853^{+0.001}_{-0.001}$   & -                   \\  &&\\
$\kappa_s$		&	$0.0527^{+0.0001}_{-0.0001}$ & -                   \\  &&\\
$q_\text{DM}$	&	$0.3685^{+0.0003}_{-0.0005}$	  & -                   \\  &&\\ 
$r_s$          & $25.30^{+0.03}_{-0.02}$ &     arcsec                  \\  && \\
$\eta_{\text{PPN}}$&$1.151^{+0.008}_{-0.004}$	& -                   \\  &&\\ 
$\text{shear}_{\text{mag}}$&$0.0230^{+0.0001}_{-0.0001}$ & -           \\  &&\\
$\text{shear}_{\phi}$&$35.95^{+0.04}_{-0.03}$& degree                \\ [1ex]
\hline
\end{tabular}
\end{table}

\begin{table}
\centering
\caption{Most probable values and $68\%$ confidence intervals for the composite model similar to \citet{Wong2015}. For the description of the parameters see Sec.~\ref{Sec. Modelling}.}
\begin{tabular}{lcc} 
\hline
 Parameter  & {MP}$_5$ & Physical Units\\
\hline 
$\Upsilon$	&   $5.437^{+0.002}_{-0.001}$	  & M$_\odot$/L$_\odot$  \\  &&\\
$i$		        &	$88.44^{+0.03}_{-0.07}$   & degree               \\  &&\\ 
$\beta_z$		&	$-0.606^{+0.004}_{-0.001}$   & -                   \\  &&\\
$\kappa_s$		&	$0.0306^{+0.0001}_{-0.0002}$ & -                   \\  &&\\
$r_s$          &  $19.19^{+0.05}_{-0.08}$ &     arcsec                  \\  && \\
$\log_{10}{\frac{M_{\text{\tiny{SMBH}}}}{M_\odot}}$& $6.8^{+0.02}_{-0.01}$ & - \\ &&  \\
$\eta_{\text{PPN}}$&$1.11^{+0.01}_{-0.02}$	  & -                   \\  &&\\ 
$\text{shear}_{\text{mag}}$&$0.0986^{+0.0001}_{-0.008}$ & -           \\  &&\\
$\text{shear}_{\phi}$&$83.52^{+0.01}_{-0.01}$& degree                \\ [1ex]
\hline
\end{tabular}
\end{table}

\begin{table}
\centering
\caption{Most probable values and $68\%$ confidence intervals for the fiducial model assuming $H_0 = 67.28$\,km\,s$^{-1}$\,Mpc$^{-1}$. For the description of the parameters see Sec.~\ref{Sec. Modelling}.}
\begin{tabular}{lcc} 
\hline
 Parameter  & {MP}$_5$ & Physical Units\\
\hline 
$\Upsilon_0$	&   $4.63^{+0.00}_{-0.08}$	  & M$_\odot$/L$_\odot$  \\  &&\\
$\delta$		&	$0.97^{+0.05}_{-0.05}$	  & arcsec$^{-1}$        \\  &&\\
$\upsilon_0$	&	$0.61^{+0.04}_{-0.03}$	  & -             \\  &&\\
$i$		        &	$71^{+3}_{-3}$   & degree               \\  &&\\ 
$\beta_z$		&	$-0.91^{+0.04}_{-0.06}$   & -                   \\  &&\\
$\kappa_s$		&	$0.102^{+0.005}_{-0.005}$ & -                   \\  &&\\
$q_\text{DM}$	&	$0.46^{+0.02}_{-0.02}$	  & -                   \\  &&\\ 
$\eta_{\text{PPN}}$&$1.05^{+0.04}_{-0.03}$	  & -                   \\  &&\\ 
$\text{shear}_{\text{mag}}$&$0.022^{+0.001}_{-0.001}$ & -           \\  &&\\
$\text{shear}_{\phi}$&$50^{+3}_{-3}$& degree                \\ [1ex]
\hline
\end{tabular}
\end{table}

\begin{table}
\centering
\caption{Most probable values and $68\%$ confidence intervals for the fiducial model assuming $H_0 = 68.20$\,km\,s$^{-1}$\,Mpc$^{-1}$. For the description of the parameters see Sec.~\ref{Sec. Modelling}.}
\begin{tabular}{lcc} 
\hline
 Parameter  & {MP}$_5$ & Physical Units\\
\hline 
$\Upsilon_0$	&   $4.46^{+0.07}_{-0.08}$	  & M$_\odot$/L$_\odot$  \\  &&\\
$\delta$		&	$1.27^{+0.07}_{-0.06}$	  & arcsec$^{-1}$        \\  &&\\
$\upsilon_0$	&	$0.88^{+0.06}_{-0.03}$	  & -            \\  &&\\
$i$		        &	$71^{+2}_{-3}$   & degree               \\  &&\\ 
$\beta_z$		&	$-0.56^{+0.03}_{-0.03}$   & -                   \\  &&\\
$\kappa_s$		&	$0.100^{+0.004}_{-0.005}$ & -                   \\  &&\\
$q_\text{DM}$	&	$0.49^{+0.03}_{-0.02}$	  & -                   \\  &&\\ 
$\eta_{\text{PPN}}$&$1.03^{+0.04}_{-0.04}$	  & -                   \\  &&\\ 
$\text{shear}_{\text{mag}}$&$0.0225^{+0.0009}_{-0.0015}$ & -           \\  &&\\
$\text{shear}_{\phi}$&$56^{+2}_{-2}$& degree                \\ [1ex]
\hline
\end{tabular}
\end{table}

\begin{table}
\centering
\caption{Most probable values and $68\%$ confidence intervals for the fiducial model assuming $H_0 = 74.2$\,km\,s$^{-1}$\,Mpc$^{-1}$. For the description of the parameters see Sec.~\ref{Sec. Modelling}.}
\begin{tabular}{lcc} 
\hline
 Parameter  & {MP}$_5$ & Physical Units\\
\hline 
$\Upsilon_0$	&   $5.22^{+0.09}_{-0.08}$	  & M$_\odot$/L$_\odot$  \\  &&\\
$\delta$		&	$1.86^{+0.07}_{-0.05}$	  & arcsec$^{-1}$        \\  &&\\
$\upsilon_0$	&	$0.94^{+0.04}_{-0.03}$	  & -             \\  &&\\
$i$		        &	$80^{+5}_{-3}$   & degree               \\  &&\\ 
$\beta_z$		&	$-0.89^{+0.03}_{-0.05}$   & -                   \\  &&\\
$\kappa_s$		&	$0.079^{+0.005}_{-0.004}$ & -                   \\  &&\\
$q_\text{DM}$	&	$0.48^{+0.02}_{-0.04}$	  & -                   \\  &&\\ 
$\eta_{\text{PPN}}$&$1.11^{+0.04}_{-0.04}$	  & -                   \\  &&\\ 
$\text{shear}_{\text{mag}}$&$0.025^{+0.002}_{-0.002}$ & -           \\  &&\\
$\text{shear}_{\phi}$&$62^{+2}_{-2}$& degree                \\ [1ex]
\hline
\end{tabular}
\end{table}

\begin{table}
\centering
\caption{Most probable values and $68\%$ confidence intervals for the fiducial model assuming $H_0 = 72.2$\,km\,s$^{-1}$\,Mpc$^{-1}$. For the description of the parameters see Sec.~\ref{Sec. Modelling}.}
\begin{tabular}{lcc} 
\hline
 Parameter  & {MP}$_5$ & Physical Units\\
\hline 
$\Upsilon_0$	&   $5.11^{+0.06}_{-0.06}$	  & M$_\odot$/L$_\odot$  \\  &&\\
$\delta$		&	$0.136^{+0.01}_{-0.008}$	  & arcsec$^{-1}$        \\  &&\\
$\upsilon_0$	&	$0.43^{+0.03}_{-0.02}$	  & -             \\  &&\\
$i$		        &	$83^{+4}_{-3}$   & degree               \\  &&\\ 
$\beta_z$		&	$-0.78^{+0.03}_{-0.04}$   & -                   \\  &&\\
$\kappa_s$		&	$0.074^{+0.003}_{-0.002}$ & -                   \\  &&\\
$q_\text{DM}$	&	$0.46^{+0.01}_{-0.02}$	  & -                   \\  &&\\ 
$\eta_{\text{PPN}}$&$1.13^{+0.02}_{-0.03}$	  & -                   \\  &&\\ 
$\text{shear}_{\text{mag}}$&$0.026^{+0.001}_{-0.001}$ & -           \\  &&\\
$\text{shear}_{\phi}$&$61^{+3}_{-2}$& degree                \\ [1ex]
\hline
\end{tabular}
\end{table}

\begin{table}
\centering
\caption{Most probable values and $68\%$ confidence intervals for the fiducial model assuming $H_0 = 76.2$\,km\,s$^{-1}$\,Mpc$^{-1}$. For the description of the parameters see Sec.~\ref{Sec. Modelling}.}
\begin{tabular}{lcc} 
\hline
 Parameter  & {MP}$_5$ & Physical Units\\
\hline 
$\Upsilon_0$	&   $5.16^{+0.08}_{-0.08}$	  & M$_\odot$/L$_\odot$  \\  &&\\
$\delta$		&	$0.28^{+0.03}_{-0.04}$	  & arcsec$^{-1}$        \\  &&\\
$\upsilon_0$	&	$0.79^{+0.06}_{-0.08}$	  & -             \\  &&\\
$i$		        &	$83^{+4}_{-3}$   & degree               \\  &&\\ 
$\beta_z$		&	$-0.44^{+0.02}_{-0.03}$   & -                   \\  &&\\
$\kappa_s$		&	$0.077^{+0.005}_{-0.003}$ & -                   \\  &&\\
$q_\text{DM}$	&	$0.52^{+0.03}_{-0.03}$	  & -                   \\  &&\\ 
$\eta_{\text{PPN}}$&$1.14^{+0.04}_{-0.04}$	  & -                   \\  &&\\ 
$\text{shear}_{\text{mag}}$&$0.028^{+0.002}_{-0.001}$ & -           \\  &&\\
$\text{shear}_{\phi}$&$65^{+2}_{-2}$& degree                \\ [1ex]
\hline
\end{tabular}
\end{table}


\bsp	
\label{lastpage}
\end{document}